\documentclass[twocolumn,showpacs,
 amsmath,amssymb,
 aps,
pra,
]{revtex4-1}

\usepackage{graphicx}
\usepackage{dcolumn}
\usepackage{bm}

\begin{document}

\preprint{APS/123-QED}

\title{Hysteretic Mutual Synchronization of PERP-STNO Pairs Analyzed by a Generalized Pendulum-like Model}

\author{Hao-Hsuan Chen}
\email{HaoHsuanChen@hotmail.com}
\affiliation{Hefei Innovation Research Institute, Beihang University, Hefei, 230013, China}

\author{Ching-Ming Lee}
\email{cmlee@yuntech.edu.tw}
\affiliation{Graduate School of Materials Science, National Yunlin University of Science and Technology, Douliou, 64002, Taiwan}

\author{Lang Zeng}%
\author{Wei-sheng Zhao}
\email{weisheng.zhao@buaa.edu.cn}
\affiliation{%
 Fert Beijing Institute, BDBC, and the School of Integrated Circuit Science and Engineering, Beihang University, Beijing 100191, China
}%

\author{Ching-Ray Chang}
\affiliation{%
Department of Physics, National Taiwan University, Taipei 10617, Taiwan
}%





\date{\today}

\begin{abstract}
At present, the Kuramoto model is the standard and widely accepted theoretical approach for analyzing the synchronization of spin-torque nano-oscillators (STNOs) coupled by an interaction. Nevertheless, the oscillatory decaying regime as well as the initial condition (IC)-dependence (hysteretic) that exist in the synchronization of many types of STNOs cannot be explained by this model. In order to more precisely elucidate the physical mechanisms behind the two phenomena, in this paper we develop a generalized pendulum-like model based on the two common features of non-linear auto-oscillators: one is the stability of the amplitude/energy of dynamic states; the other is the non-linear dynamic state energy of oscillators. In this new model, we find that the Newtonian-like particle with sufficient kinetic energy can overcome the barrier of phase-locking potential to evolve into a stable asynchronization (AS) state, leading to the (IC)-dependent synchronization. Furthermore, due to the presence of the kinetic energy, this particle can also oscillate around the minima of the phase-locking potential, leading to the oscillatory decaying regime. Thereby, in this work, we adopt this new model to analyze the IC-dependent mutual synchronization of perpendicular-to-plane (PERP)-STNO pairs, and then we suggest that the initial conditions can be controlled to avoid such a phenomenon by using magnetic dipolar coupling.
\end{abstract}

\maketitle


\section{\label{sec:level1}INTRODUCTION}
Spin-Transfer Torque (STT)\cite{slonczewski1996current,berger1996emission,Slonczewski2002,XiaoJiang2004,Hirsch1999}  can, as a negative damping effect, maintain persistent magnetic auto-oscillations in the GHz to sub-THz frequency range by continuously injecting energy into the magnetic system so as to resist its energy dissipation. Such auto-oscillators are described as Spin-Torque Nano-Oscillators (STNOs). So far, several kinds of STNOs have been reported, including STNOs based on the quasi-uniform mode in nano-pillars (NPs) \cite{kiselev2003microwave,Houssameddine2007,Kubota2013}, nano-contacts (NCs) \cite{kaka2005}, non-uniform magnetic solitons \cite{Pribiag2007,Khvalkovskiy2009,Hoefer2010,XiaoD2017,Garcia-Sanchez2016}, and anti-ferromagnetism
\cite{Cheng2016,Khymyn2017,Shen2019}. Owing to  promising applications as microwave radiation sources\cite{kiselev2003microwave}, communication devices\cite{Consolo2010,Choi2014}, as well as applications in neuromorphic computation\cite{Romera2018}, STNOs have become an emerging research topic in the field of spintronics.\\

However, practical application of STNOs is limited by issues such as low emitted power and large linewidth. A feasible approach to overcome these drawbacks has been to synchronize an array of multiple STNOs via some coupling mechanism. So far, several types of coupling mechanisms have been reported namely pioneering propagating spin waves based on NC structure \cite{kaka2005,Mancoff2005}; existence of electric coupling in the circuit based on NP structure \cite{grollier2006,Taniguchi2018}; magnetic dipolar coupling based on NPs structure (quasi-uniform or vortex modes) \cite{HaoHsuan2011,HaoHsuan2012,Belanovsky2012,AbreuAraujo2015,HaoHsuan2016,Kang2018,HaoHsuan2018,Mancilla-Almonacid2019,Li2017}, NCs structure (droplets) \cite{Wang2017,Wang2018a}, and nano-constriction NC structure driven by the spin Hall effect (SHE) \cite{Awad2016,Awad2018}. Among these schemes, the last scheme proposed by Ref. \cite{Awad2018} has so far achieved the highest recorded number of synchronized STNOs in experiments. Thus, we believe that magnetic dipolar coupling is a more promising synchronization scheme, such as synchronized perpendicular-to-plane polarizer (PERP)-STNO pairs by magnetic dipolar coupling, which does not need the assistance of an external field and can be driven by opposite currents\cite{HaoHsuan2016,HaoHsuan2018}.

Previously we have adopted an age-old pendulum-like model to theoretically solve the initial condition (IC)-dependent excitation, namely, parallel (P)/anti-parallel (AP)/OP coexistent states in an individual PERP-STNO applied by an external field normal to the film plane\cite{HaoHsuan2017}, which cannot be explained by the Kuramoto model\cite{Acebron2005}. Here, for the initial states with less kinetic energy, the Newtonian-like particle (magnetization) will decay in an oscillating manner into the P or AP states, while for the initial states with sufficient kinetic energy, the magnetization might be excited into the OP precessional states for a small enough damping constant. The previous model is developed based on a very narrow assumption of a strong in-plane shape anisotropy, i.e. demagnetization energy, thereby it can only be used to analyze the OP precessional states with the lower energy level or in the lower current exciting regime. Moreover, the previous model is unsuitable for other types of STNOs without a strong demagnetization energy, such as perpendicular magnetized anisotropy (PMA)-STNOs \cite{Taniguchi_2013}. Even with these drawbacks, the previous model has been successfully adopted to well analyze the mutual synchronization of PERP-STNO pairs in the lower current regime\cite{HaoHsuan2011,HaoHsuan2016,HaoHsuan2018}, where the oscillatory decaying regime has been observed. This implies that the previous model should also be able to solve the existence of IC-dependent mutual synchronization of PERP-STNOs, namely, phase-locking (PL)/asynchronization (AS) coexistent state. In addition, such a phenomenon has been observed and analyzed in the pioneering works of the injection-locking of other types of STNOs\cite{Bonin2009,Tabor2010,Zhou2010,Li2010,DongLi2011,DAquino2017,Tortarolo2018}, indicating that the phenomenon should be the common feature of all kinds of STNOs and thereby can be explained by the pendulum model. In reality, all types of STNOs have two common characteristics of non-linear auto-oscillators: one is the stability of dynamic states, which ensures the persistent oscillation of STNOs; the other is the non-linear dynamic state energy, in which the frequency of STNOs depends on amplitude/energy. Therefore, in order to precisely elucidate the physical mechanisms behind the IC-dependent synchronization, it is necessary to derived a generalized pendulum-like model for all kinds of STNOs based on these two common features.

In this paper, we aim to develop a generalized pendulum-like model to intuitively unveil the mechanisms of IC-dependent mutual synchronization of coupled non-linear auto-oscillators. The paper is organized as follows: In section \ref{a}, we first develop a new theoretical approach, termed the \textit{local} coordinate transformation, to determine the stability for two-dimensional auto-oscillatory states in a more straightforward way. In section \ref{B} and Appendix \ref{appa}, based on the local coordinate transformation, the generalized pendulum-like coupled equations are straightforwardly derived for coupled non-linear auto-oscillators with weak interactions. Subsequently, we analyze an individual forced pendulum phase-locked by an oscillating uniform gravity as an instructive example of synchronizing non-linear auto-oscillators. In this model, IC-dependent phase-locking as well as oscillatory decaying transient state are easily explained. In section \ref{C}, we give the theoretical framework, i.e. the pendulum-like model, to analyze the mutual synchronization of a pair of PERP-STNOs with magnetic dipolar coupling, which are connected in parallel and in serial, respectively. In section \ref{D}, the phase diagrams for the mutual synchronization of PERP-STNO pairs, hysteretic frequency responses, locking phase angles, and transient state of the synchronization are all obtained from the theoretical model. Meanwhile, we perform the pendulum-like model and macrospin simulations to verify our analytical results. Finally, a brief summary and discussions about how to avoid the IC-dependent synchronization are given in section \ref{sec:3}.

\section{\label{sec:2}Model and Theory}
\subsection{\label{a}Generalized local Coordinate Transformation}
\begin{figure*}
\begin{center}
\includegraphics[width=12.5cm]{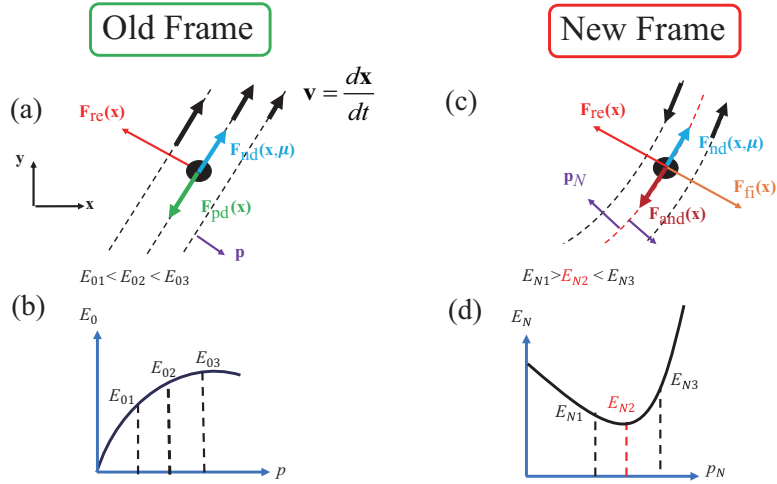}
\end{center}
\caption{(Color online) Illustration of local coordinate transformation in the viewpoint of the force/field. Here (a) and (c) present the motions of an auto-oscillatory system in the old (O) and new (N) frames, respectively. The family of dashed curves in (a) and (c) denote the trajectories given by the effective energies $E_{0}$ and $E_{N}$, respectively. The red dashed curve in (b) is a meta-stable state. The black, red, blue, green, and purple arrows in (a) denote phase velocity $\mathbf{v}$, restoring force/field $\mathbf{F}_{\mathrm{re}}(\mathbf{x})$, positive damping force/field $\mathbf{F}_{\mathrm{pd}}(\mathbf{x})$, negative damping force/field $\mathbf{F}_{\mathrm{nd}}(\mathbf{x},\mu)$, and unit vector $\mathbf{p}$, respectively. The orange, brown, and purple arrows in (c) are the fictitious force/field $\mathbf{F}_{\mathrm{fi}}(\mathbf{x})$, anti-negative damping force/field $\mathbf{F}_{\mathrm{and}}(\mathbf{x})$, and unit vector $\mathbf{p}_{N}$, respectively. Figures (b) and (d) present the effective energies as a function of variables $p$ and $p_{N}$, respectively. }
\label{Forceevrsion}
\end{figure*}

 For a two-dimensional auto-oscillator, the stability is a vital and fundamental issue, from which one can find out the physical conditions to maintain persistent oscillatory states qualitatively and quantitatively. For this reason, we need to develop a new theoretical viewpoint and approach that is independent of coordinate system in order to determine the stability of two-dimensional auto-oscillators in a more straightforward way. Another reason we develop this new approach is to quantitatively estimate the perturbation level of auto-oscillatory trajectories caused by weak interactions among multiple auto-oscillators, and therefore obtain the equations of the phase angle dynamics for multiple coupled auto-oscillators (see Appendix \ref{appa}). This cannot be directly achieved from the approach based on the energy-averaging technique\cite{GBertotti2009nonlinear}.

The general vector form for a two-dimensional autonomous system can be expressed as \cite{GBertotti2009nonlinear}
\setlength\abovedisplayskip{6pt}
\setlength\belowdisplayskip{6pt}
\begin{eqnarray}
\frac{d\textbf{x}}{dt}=\textbf{f}
(\textbf{x}).
\label{vectform}
\end{eqnarray}
Here, $\mathbf{x}$ indicates the state vector of the system on a two dimensional phase plane. According to the Helmholz decomposition theorem\cite{GBertotti2009nonlinear}, the tangent vector field  $\textbf{f}(\textbf{x})$ on the phase plane can be uniquely decomposed into two components: one is the \textit{divergence free} term $\textbf{f}_{\mathrm{d}}(\textbf{x})=(-\nabla_{\mathbf{x}}
E_{0})\times\mathbf{n}$; the other is \textit{curl free} term $\textbf{f}_{\mathrm{c}}(\textbf{x})=-\nabla_{\mathbf{x}}\Omega$.
Here, $E_{0}(\mathbf{x})$, $\Omega(\mathbf{x})$, and $\mathbf{n}$ are scalar potentials and the unit vector normal to the plane, respectively. Furthermore, according to the balance equation of $E_{0}(\mathbf{x})$, only the component of $\textbf{f}_{\mathrm{c}}(\textbf{x})$ normal to $\textbf{f}_{\mathrm{d}}(\textbf{x})$ can contribute to the time rate of $E_{0}(\mathbf{x})$. Also, if $E_{0}(\mathbf{x})$ and $\textbf{f}_{\mathrm{c}}(\textbf{x})$ can be treated as energy and damping terms, respectively, then the component of $\textbf{f}_{\mathrm{c}}(\textbf{x})$ along $\textbf{f}_{\mathrm{d}}(\textbf{x})$ is sufficiently weak to hardly affect the oscillatory states governed by $\textbf{f}_{\mathrm{d}}(\textbf{x})$ through microscopic interactions with the environment.
 Thus, it is quite reasonable to assume $\textbf{f}_{\mathrm{d}}(\textbf{x})$ and $\textbf{f}_{\mathrm{c}}(\textbf{x})$ are normal to each other everywhere on the phase plane\cite{GBertotti2009nonlinear}, that is, $\textbf{f}_{\mathrm{c}}(\mathbf{x})\approx-\Gamma(\mathbf{x})
 \nabla_{\mathbf{x}}E_{0}$. Here, $\Gamma(\mathbf{x})$ is the damping rate, which contains the positive and negative damping effects. \\

Therefore, Eq. (\ref{vectform}) can be further written as
\setlength\abovedisplayskip{6pt}
\setlength\belowdisplayskip{6pt}
\begin{eqnarray}
\frac{d\textbf{x}}{dt}&=&-(\nabla_{\mathbf{x}} E_{0})\times\mathbf{n}+\left(-\Gamma(\mathbf{x})\right)
\nabla_{\mathbf{x}} E_{0}
,\nonumber\\
&=&-(\nabla_{\mathbf{x}} E_{0})\times\mathbf{n}+\left[-\alpha(\mathbf{x})
\frac{d\mathbf{x}}{dt}+
a(\mathbf{x},\mu)(\mathbf{n}\times\mathbf{p})\right]\nonumber\\
&&\times\mathbf{n}.
\label{vectformappro}
\end{eqnarray}
The first term on the right side of Eq. (\ref{vectformappro}) defines a group of energy conserved trajectories(states) $C(E_{0})$ designated by $E_{0}$, implying that the effective particle is moving along
the direction normal to
$-\nabla_{\mathbf{x}} E_{0}(\mathbf{x})$.
Besides, the first and second expressions in Eq. (\ref{vectformappro}) are both equivalent. The damping term in the second expression is derived from the \textit{Rayleigh dissipation function}. The unit vector $\mathbf{p}\equiv\nabla_{\mathbf{x}}
E_{0}/|\nabla_{\mathbf{x}}E_{0}|$ is normal to the conservative trajectories. In the second expression, the non-conservative part contains two terms: first one is the positive damping term with a \textit{positive} damping factor $\alpha(\mathbf{x})>0$; second one is the negative damping term with a \textit{negative} damping factor $a(\mathbf{x},\mu)$, in which $\mu$ is used to adjust its intensity.\\

Actually, Eq. (\ref{vectformappro}) can also be equivalently expressed in the viewpoint of force/field\cite{Thiele1973}:
\begin{eqnarray}
\mathbf{n}\times\frac{d\textbf{x}}{dt}
&=&\mathbf{F}_{\mathrm{re}}(\mathbf{x})+\mathbf{F}_{\mathrm{pd}}
(\mathbf{x})+\mathbf{F}_{\mathrm{nd}}
(\mathbf{x}).
\label{forceversion}
\end{eqnarray}
The left-hand side of Eq. (\ref{forceversion}) describes the gyro-force/field. The three forces/fields on the right side of Eq. (\ref{forceversion}) are given in turn as: restoring force/field $\mathbf{F}_{\mathrm{re}}(\mathbf{x})=-\nabla_{\mathbf{x}}E_{0}$ ; positive damping force/field $\mathbf{F}_{\mathrm{pd}}(\mathbf{x})=-\alpha(\mathbf{x})( d\mathbf{x}/dt)$; negative damping force/field $\mathbf{F}_{\mathrm{nd}}(\mathbf{x})=a(\mathbf{x},\mu)
(\mathbf{p}
\times\mathbf{n})$. As illustrated in Fig. \ref{Forceevrsion} (a) and (b), when $\mathbf{F}_{\mathrm{pd}}(\mathbf{x})$ and $\mathbf{F}_{\mathrm{nd}}(\mathbf{x})$ compensate for each other, the dynamic states satisfying $\partial E_{0}/\partial p\neq0$ (see Fig. \ref{Forceevrsion}(b)) can be driven by $\mathbf{F}_{\mathrm{nd}}(\mathbf{x,\mu})$. Here, the state vector can be expressed as $\mathbf{x}=p(\mathbf{x})\mathbf{p}+\phi(\mathbf{x})
\mathbf{\hat\phi}$, and $\mathbf{\hat\phi}\equiv\mathbf{n}\times\mathbf{p}$.\\

 As illustrated in Fig. \ref{Forceevrsion}, to confirm the stability of the dynamic states, one can use the following transformation to ensure that the moving effective particle is stationary in the new frame:
\setlength\abovedisplayskip{6pt}
\setlength\belowdisplayskip{6pt}
\begin{eqnarray}
\left[\frac{d\mathbf{x}}{dt}\right]_{\mathrm{N}}&=&\left[
\frac{d\mathbf{x}}{dt}\right]
_{\mathrm{O}}-\mathbf{v}_{p}(\mathbf{x}), \nonumber\\
&=&\left[\frac{d\mathbf{x}}{dt}\right]
_{\mathrm{O}}-\left[v_{p}(\mathbf{x})\mathbf{p}\right]
\times\mathbf{n},
\label{RotTransgeneral}
\end{eqnarray}
Here, the abbreviations "N" and "O" denote the new and old frames, respectively.
Besides, $v_{p}(\mathbf{x})$ is a scalar velocity field, so here Eq. (\ref{RotTransgeneral}) is termed as a \textit{local} coordinate transformation. By using Eq. (\ref{RotTransgeneral}), there are two new forces/fields induced in the new frame (see Fig. \ref{Forceevrsion}(c)): one is the fictitious force/field $\mathbf{F}_{\mathrm{fi}}(\mathbf{x})=-
v_{p}(\mathbf{x})\mathbf{p}$; the other is the anti-negative damping force/field $\mathbf{F}_{\mathrm{and}}(\mathbf{x})=-\alpha(\mathbf{x})v_{p}
(\mathbf{x})(\mathbf{p}\times\mathbf{n})$. If $\mathbf{v}_{p}$ is chosen so that $\mathbf{F}_{\mathrm{and}}$ cancels $\mathbf{F}_{\mathrm{nd}}$ on the phase plane,
Eq. (\ref{forceversion}) will take the following form in the new frame:
\setlength\abovedisplayskip{6pt}
\setlength\belowdisplayskip{6pt}
\begin{eqnarray}
\mathbf{n}\times\left[\frac{d\textbf{x}}{dt}\right]_{\mathrm{N}}
&=&\mathbf{F}_{\mathrm{N,re}}(\mathbf{x})
+\mathbf{F}_{\mathrm{N,pd}}
(\mathbf{x}).
\label{forceversionN}
\end{eqnarray}

Here, $\mathbf{F}_{\mathrm{N,re}}(\mathbf{x})=-\nabla_{\mathbf{x}}E_{N}$,
$E_{N}(\mathbf{x})=E_{0}(\mathbf{x})+U'_{N}(\mathbf{x})$, and
$U'_{N}(\mathbf{x})\equiv\int^{\mathbf{x}} d\mathbf{x}'\cdot[v_{p}(\mathbf{x}')\mathbf{p}]$. Also, $\mathbf{F}_{\mathrm{N,pd}}(\mathbf{x})
=-\alpha(\mathbf{x})(d\mathbf{x}/dt)_{\mathrm{N}}$, and $\alpha(\mathbf{x})v_{p}(\mathbf{x})=a(\mathbf{x},\mu)$ has been utilized.\\

In principle, for stable auto-oscillatory states, the contour of $E_{N}$ exhibits canyon-like shapes, i.e. stable limit cycles. At anywhere on the bottom of the canyon, the values of $E_{N}$ are all equal, implying that the effective particle can be static at anywhere on the bottom of the canyon.
For these canyons, one can determine them by walking along $\mathbf{p}_{N}=\nabla_{\mathbf{x}}E_{N}/|\nabla_{\mathbf{x}}E_{N}|$. And then, since the state vector in the new frame can be expanded as $\mathbf{x}=p_{N}(\mathbf{x})\mathbf{p}_{N}+\phi_{N}(\mathbf{x})
\mathbf{\hat\phi}_{N}$, where $\mathbf{\hat\phi}_{N}\equiv\mathbf{n}\times\mathbf{p}_{N}$, $E_{N}$ must be only a function of $p_{N}(\mathbf{x})$. Thus, by requiring $(\partial E_{N}/\partial P_{N})_{P_{N0}}=0$ as well as $(\partial^{2}E_{N}/\partial P_{N}^{2})_{P_{N0}}>0$, we can find out the positions of
these canyons and confirm their stability exactly, as illustrated in Fig. \ref{Forceevrsion}(d).\\

Strictly speaking, because $\mathbf{v}_{p}(\mathbf{x})$ in Eq. (\ref{RotTransgeneral}) depends not only on $p$ but also on $\phi$ (see also Appendix \ref{app:1}), the actual auto-oscillatory trajectories, which are the meta-stable states of $E_{N}(\mathbf{x})$, are not exactly the same as the dynamic ones of $E_{0}(\mathbf{x})$, as illustrated in Figs. \ref{Forceevrsion} (a) and (c). However, for most cases of auto-oscillators, due to the much larger order of $|\mathbf{F}_{\mathrm{re}}(\mathbf{x})|$ than those of $|\mathbf{F}_{\mathrm{pd}}(\mathbf{x})|$ and $|\mathbf{F}_{\mathrm{nd}}(\mathbf{x},\mu)|$, the
auto-oscillations can be roughly treated as the dynamic states of $E_{0}(\mathbf{x})$, that is, $p_{N}(\mathbf{x})\approx p(\mathbf{x})$.

In the following, we take two simple examples, namely PERP-STNOs\cite{Chen2019} and PMA-STNOs\cite{Taniguchi_2013}, to briefly introduce the practical usage of this new theory in analyzing the stability of STNOs. First, since these two types of STNOs both have axial symmetric anisotropic energies, their dynamic states (out-of-plane (OP) precessions) can both be designated by the z-component of the magnetization, i.e. $m_{z}$. Subsequently, their the dynamic state energies can be expressed as $E_{0}=(-k/2)m^{2}_{z}$, where $k<0$ in PERP-STNOs and $k>0$ in PMA-STNOs. Interestingly, notice that in PERP-STNOs the dynamic state energy has a positive curvature $\partial^{2}E_{0}/\partial m_{z}^{2}=-k>0$; while in PMA-STNOs the curvature is negative, $\partial^{2}E_{0}/\partial m_{z}^{2}=-k<0$. From the viewpoint of $E_{N}(m_{z})$, this implies that the dynamic state energy of PERP-STNOs itself possesses the potential to produce the stability of OP precessions. The effective energy $U'_{N}(m_{z})$ produced by the STT was proven to just shift the equilibrium OP precessional point $m_{z0}$ away from the film plane, instead of changing the curvature of $E_{0}(m_{z})$ (see Ref.\cite{Chen2019}). Thus, PERP-STNOs can be excited solely by currents with opposite injection directions without the assistance of an external magnetic field.

 However, in PMA-STNOs we need a different type of STT. It has been proven that this can be provided by an in-plane polarizer with an asymmetric spin polarization factor, in order to turn the negative curvature of $E_{0}$ into a positive one of $E_{N}(m_{z})$(see Ref.\cite{Taniguchi_2013}). Thereby, PMA-STNOs can be excited only by positive current. Notably, due to the non-axially symmetric polarizer in PMA-STNOs, one can calculate $U'_{N}(m_{z})$ by using the approach introduced in Appendix \ref{app:1}. Furthermore, since the STT fails to shift the equilibrium point $m_{z0}$ of the OP state in PMA-STNOs, it is necessary to apply an external field along the z-axis to shift $m_{z0}$ away from the film plane. Here, the zeeman energy produced by the external field does not change the curvature of $E_{N}(m_{z0})$.

 From the brief analyses above, we can conclude that for different types of two-dimensional dynamic state energies, the physical conditions to excite auto-oscillatory states are also different. In principle, all types of dynamic state energies have the potential to excite auto-oscillatory states, including (anti-)ferromagnetic exchange coupling, shape anisotropic energy, perpendicular magnetic anisotropic energy, zeeman energy, etc. However, in nature there are only a limited number of negative damping forces that exist, e.g. STT, and so dynamic states of some type do not remain stable. Finally,
 due to the positive curvature of $E_{N}(\mathrm{x})$ for all types of stable auto-oscillatory states, one can straightforwardly obtain the governing equation of phase angle for coupled multiple oscillators, as described in the following section.

\subsection{\label{B}Generalized Pendulum-like Model}
\begin{figure*}
\begin{center}
\includegraphics[width=12cm]{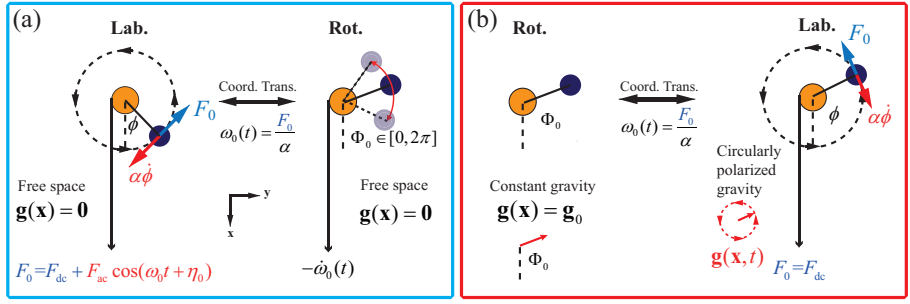}
\end{center}
\caption{(Color online) Phase-locking of a non-linear auto-oscillator vividly interpreted by a forced pendulum in the laboratory (Lab.) and rotating (Rot.) frames, respectively. Here, the rotating frame is rotating at a velocity $\omega_{0}(t)=F_{0}/\alpha$. (a)\textbf{Lab}: Pendulum in free space is driven by a force  $F_{0}=F_{\mathrm{dc}}+F_{\mathrm{ac}}\cos(\omega_{0}
t+\eta_{0})$. \textbf{Rot}: Pendulum in free space is driven by an effective force  $-\dot{\omega_{0}}(t)=(F_{\textrm{ac}}/\alpha)\sin(\omega_{0}t+\eta_{0})$. (b) \textbf{Rot}: Pendulum is placed in a constant gravity $\textbf{g}(\textbf{x})=\textbf{g}_{0}$, which has an angle $\Phi_{0}$ relative to the x-axis. \textbf{Lab}: Pendulum driven by a dc force $F_{\mathrm{dc}}$ is subject to a circularly polarized oscillating gravitational field $\textbf{g}(\textbf{x},t)=g_{0}[\textbf{x}\cos(\omega_{e}t+\Phi_{0})+\textbf{y}\sin(\omega_{e}
t+\Phi_{0})]$. Here, $\omega_{e}=\omega_{0}$.}
\label{Pendulum}
\end{figure*}
For auto-oscillators with a strong \textit{nonlinear frequency shift coefficient} $(\partial^{2} E_{0}/\partial p^{2})_{p_{0}}$ (see Fig. \ref{Forceevrsion}(b)as an example), the dynamics of phase angle $\phi$ will be coupled with that of momentum $p$\cite{Slavin2009,Zhou2010}, as detailed in the Appendix \ref{app:2}. Based on this finding, the generalized pendulum-like particles for coupled auto-oscillators by weak interactions can be straightforwardly derived through the local coordinate transformation as (see Appendix \ref{app:2}):
\setlength\abovedisplayskip{6pt}
\setlength\belowdisplayskip{6pt}
\begin{eqnarray}
\ddot{\phi_{i}}+\alpha_{\mathrm{eff},i}(p_{i0})\dot{\phi_{i}}&=&
F_{\mathrm{eff},i}(p_{i0})+\alpha_{\mathrm{eff},i}(p_{i0})\nonumber\\
&&\times\sum_{j=1(j\neq i)}^{n}\left(\frac{\partial U_{I}}{\partial p_{i}}\right)_{p_{i0}}\nonumber\\
&&-H^{(2)}_{Oi}(p_{i0})\sum_{j=1(j\neq i)}^{n}\left(\frac{\partial U_{I}}{\partial\phi_{i}}\right)_{p_{i0}}\nonumber\\
&&+\sum_{l=1}^{n}\frac{\partial}{\partial\phi_{l}}\left(\frac{\partial U_{I}}{\partial p_{i}}\right)_{p_{i0}}\dot{\phi}_{l},
\label{nonconpenduL}
\end{eqnarray}
where $\alpha_{\mathrm{eff},i}(p_{i0})\equiv\alpha_{i}S_{i}(p_{i0})H^{(2)}_{N,i0}(p_{i0})$ and $F_{\mathrm{eff},i}(p_{i0})\equiv H_{N,i0}^{(2)}(p_{i0})
a_{i}(p_{i0},\mu)$ are the effective damping constants and effective driving forces, respectively. Also, $H^{(2)}_{N,i0}(p_{i0})$, $S_{i}(p_{i0})$, $a_{i}(p_{i0},\mu_{i})$, and $H^{(2)}_{Oi}(p_{i0})$ are the stability, positive damping function, negative damping function, and nonlinear frequency shift coefficient \cite{Slavin2009} of auto-oscillators, respectively, which are defined in Appendix \ref{app:2}.
$p_{i0}$ and $\phi_{i}$ indicate the stable equilibrium generalized canonical momenta and coordinates, respectively, which are defined in Appendix \ref{app:1}. $U_{I}$ denote weak interactions. The details of the derivation can be found in Appendix \ref{app:2}.

To obtain insight into the the synchronization of coupled auto-oscillators, it is very instructive
to understand how to phase-lock a driven pendulum by a tangent force placed in a free space, as illustrated in Fig. \ref{Pendulum}. An individual driven pendulum actually contains all four ingredients of a typical two-dimensional \textit{non-linear} auto-oscillator\cite{Slavin2009}(see also Appendix \ref{app:2}), including a \textit{non-linear} dynamic state energy  (kinetic energy) associated with the amplitude of angular (phase) velocity $|\dot{\phi}|$, positive damping (friction force $-\alpha\dot{\phi}$), negative damping (dc driving force $F_{\mathrm{dc}}$), and stability of oscillatory states (positive ratio of $\alpha/m$). Here, both the pendulum mass $m$ and the rob length $l$ has been normalized to one. Once $\alpha\dot{\phi}$ and $F_{\mathrm{dc}}$ come to balance, the pendulum will rotate permanently around the pivot with a terminal angular velocity $\dot{\phi}_{T}=F_{\mathrm{dc}}/\alpha$ without the assistance of the ac driving force $F_{\mathrm{ac}}$. Such a terminal velocity motion just reflects the nature of non-linear auto-oscillators, where the angular velocity (frequency) is strongly coupled with the momentum (amplitude or dynamic energy). This feature is very different from that of \textit{quasi-linear} auto-oscillators with a very small non-linear frequency shift coefficient\cite{Slavin2009}, as the frequency is hardly affected by the momentum or positive (negative) damping force (see also Appendix \ref{app:2})\cite{Adler1973}.

Notably, the fourth ingredient is an indispensable feature for non-linear auto-oscillators, which can be easily seen from the transient process of angular velocity: $\dot{\phi}(\tau)=Ce^{-(\alpha/m)\tau}-F_{\mathrm{dc}}/\alpha$. This means only a positive $\alpha/m$ can ensure that a velocity value slightly away from $\dot{\phi}_{T}$ will return to $\dot{\phi}_{T}$. This point satisfies with the stability of auto-oscillators defined by the local coordinate transformation.

As illustrated in the left and right panels of Fig. \ref{Pendulum}(a), a forced pendulum fails to be phase-locked by adding a small ac component $F_{\mathrm{ac}}(t)$ to $F_{\mathrm{dc}}$. This can be easily seen for such a case in the rotating frame, whose angular velocity is $\omega_{0}(t)=F_{0}/\alpha$, where the pendulum can oscillate around anywhere with $\Phi_{0}\in[0,2\pi]$ without al binding force to trap it (see the right panel of Fig \ref{Pendulum}(a)). That means some kind of anisotropic coupling force as a function of $\Phi$ is needed here to trap the particle, e.g. a uniform gravitational field/force, which is illustrated by the left panel of Fig. \ref{Pendulum}(b).

Interestingly, when transforming back to the laboratory frame, the constant gravity will be turned into a circularly polarized oscillating gravity (see the right panel of Fig. \ref{Pendulum}(b)), whose angular velocity $\omega_{e}$ is close to that of the pendulum, i.e. $\omega_{e}\approx F_{\mathrm{dc}}/\alpha$.  Then, the equation of motion for this case takes the following form:
\setlength\abovedisplayskip{6pt}
\setlength\belowdisplayskip{6pt}
\begin{eqnarray}
\ddot{\phi}+\alpha\dot{\phi}=F_{\mathrm{dc}}-g_{0}\sin \left(\phi-\omega_{e}t-\Phi_{0}\right),
\label{pendulobygra}
\end{eqnarray}
where $\Phi_{0}$ is the initial phase angle of the oscillating gravity.
Thus, it can be concluded that, in order to phase-lock a free-running non-linear auto-oscillator, it is necessary to introduce an effective anisotropic force as a function of phase angle difference ($\Phi\equiv\phi-\omega_{e}t$). In other words, if the ac driving force shown in Fig. \ref{Pendulum}(a) can be replaced with one of the form $F_{\mathrm{ac}}\cos(\phi-\omega_{e}t+\eta_{0})$ by a certain mechanism, then the phase-locking can be achieved as well. \\

In terms of $\Phi$, Eq. (\ref{pendulobygra}) becomes
\setlength\abovedisplayskip{6pt}
\setlength\belowdisplayskip{6pt}
\begin{eqnarray}
\ddot{\Phi}+\alpha\dot{\Phi}=\Delta F-g_{0}\sin \left(\Phi-\Phi_{0}\right),
\label{pendulobygrb}
\end{eqnarray}
where $\Delta F=F_{\mathrm{dc}}-\alpha\omega_{e}$.
\begin{figure}
\begin{center}
\includegraphics[width=5cm]{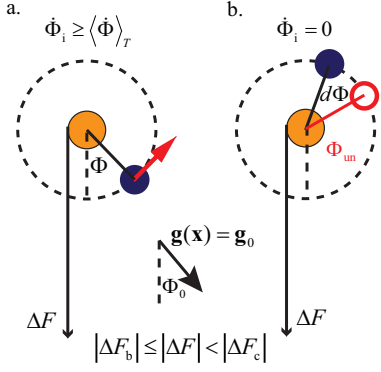}
\end{center}
\caption{(Color online) Examples of initial states evolving with certainty into AS states for $|\Delta F_{\mathrm{b}}|\leq|\Delta F|<|\Delta F_{\mathrm{c}}|$. a and b indicate the cases for
 $(\Phi_{\mathrm{i}},\dot{\Phi}_{\mathrm{i}}\geq\langle\dot{\Phi}\rangle_{T})$ and $(\Phi_{\mathrm{i}},\dot{\Phi}_{\mathrm{i}})=(\Phi_{\mathrm{un}}+d\Phi,0)$, respectively. Here, $d\Phi>0$. }
\label{initialstate}
\end{figure}
In the absence of $\alpha$ and $\Delta F$, there are two kinds of dynamic modes produced by the locking potential, i.e. uniform gravity:
one is local oscillation with energy $E<g_{0}$ around the minimum of the potential; the other is rotation
around the pivot with energy $E>g_{0}$. For the local oscillations, $\Delta F$ must be
treated as a \textit{conservative-like} force\cite{HaoHsuan2017}, where the effective potential becomes $U_{\mathrm{eff}}(\Phi)=-g_{0}\cos(\Phi-\Phi_{0})-\Delta F\Phi$. By requiring $dU_{\mathrm{eff}}/d\Phi=0$ and $d^{2}U_{\mathrm{eff}}/d\Phi^{2}>0$, the unstable and stable equilibrium points are $\Phi_{\mathrm{un}}=\Phi_{0}+\pi-\sin^{-1}(\Delta F/g_{0})$ and $\Phi_{\mathrm{PL}}=\Phi_{0}+\sin^{-1}(\Delta F/g_{0})$, respectively. Thus, the stable phase-locking (PL) state $\Phi_{\mathrm{PL}}$ of Eq. (\ref{pendulobygrb}) occurs when $|\Delta F|<g_{0}$, and $\Phi_{\mathrm{PL}}$ corresponds to the minimum of $U_{\mathrm{eff}}(\Phi)$. Obviously, there exists a upper limit value of $\Delta F$:
\setlength\abovedisplayskip{6pt}
\setlength\belowdisplayskip{6pt}
\begin{eqnarray}
|\Delta F_{\mathrm{c}}|&=&g_{0},
\label{PLDF}
\end{eqnarray}
below which the existence of PL state can be assured.

For the rotations, i.e. asynchronized (AS) states, $\Delta F$ must be considered as a \textit{non-conservative} force \cite{HaoHsuan2017}, that is, a negative damping force. Thereby, one can use the theoretical approach developed in \ref{a}
to analyze their stability. First, due to the presence of anisotropic potential, one can choose thel energy $E=\dot{\Phi}^{2}/2-g_{0}\cos(\Phi-\Phi_{0})$ as a canonical momentum $p$. By following the approach given in
appendix \ref{app:1}, one can easily obtain
\setlength\abovedisplayskip{6pt}
\setlength\belowdisplayskip{6pt}
\begin{eqnarray}
\dot{p}&\approx&-\left[\frac{\alpha}{T(p)}\right] \int_{0}^{2\pi}d\Phi\sqrt{2p+g_{0}\cos(\Phi-\Phi_{0})}\nonumber\\
&&(\mp)\left[\frac{2\pi}{T(p)}\right]\Delta F,\nonumber\\
\dot{\Phi}'&=&\frac{2\pi}{T(p)}=\frac{\partial H_{O}}{\partial p}.\nonumber
\end{eqnarray}
Here, $T(p)=\int_{0}^{2\pi}d\Phi/\sqrt{2p+2g_{0}\cos(\Phi-\Phi_{0})}$ and $H_{O}(p)=2\pi\int^{p}dp'[T(p')]^{-1}$.
 Compared to Eq. (\ref{apppphi}), the positive and damping functions are $S(p)=(1/2\pi)\int_{0}^{2\pi}d\Phi\sqrt{2p
 +2g_{0}\cos(\Phi-\Phi_{0})}$ and $a(p,\Delta F)=(\mp)[2\pi/T(p)]\Delta F$, respectively. Accordingly, the Hamiltonian
 in the new frame is: $H_{N}(p)=H_{O}(p)-\int^{p}dp'v_{p}(p')$, where $v(p)=a(p,\Delta F)/[\alpha S(p)]$. By requiring
 $\partial H_{N}/\partial p=0$ and $\partial^{2} H_{N}/\partial p^{2}>0$, one can easily obtain the criteria for stable
 AS states labeled by $p_{0}$:
\setlength\abovedisplayskip{6pt}
\setlength\belowdisplayskip{6pt}
\begin{eqnarray}
\mp\Delta F&=&\alpha S(p_{0}),\nonumber\\
\left(\frac{dS(p)}{dp}\right)_{p_{0}}&=&\int_{0}^{2\pi}d\Phi\frac{p_{0}}{\sqrt{2p_{0}+2g_{0}
\cos(\Phi-\Phi_{0})}}>0,
\nonumber\\
\label{criterAS}
\end{eqnarray}
Since $\alpha S(p_{0})>0$, a minus sign must be applied to the left-hand side of the first equation for negative $\Delta F$. In addition,
the second equation implies that equilibrium AS states ($p_{0}>g_{0}$) must be stable. For $p_{0}$ close to the energy minimum of an AS state, i.e. $E=g_{0}$, one can easily estimate the threshold of $\Delta F$ needed to drive the AS state:
\setlength\abovedisplayskip{6pt}
\setlength\belowdisplayskip{6pt}
\begin{eqnarray}
|\Delta F_{\mathrm{b}}|&=&\alpha S(p_{0}\rightarrow g_{0}),\nonumber\\
&=&\frac{\alpha\sqrt{2g_{0}}}{2\pi}S',
\label{lowerDF}
\end{eqnarray}
where $S'=\int_{0}^{2\pi}d\Phi\sqrt{1+\cos(\Phi-\Phi_{0})}\approx5.6569$.

Note that Eq. (\ref{lowerDF}) is valid for $|\Delta E|=|\int_{0}^{2\pi}d\Phi\Delta F_{\mathrm{b}}|=2\pi|\Delta F_{\mathrm{b}}|<g_{0}$ (see appendix \ref{app:1}), implying that the order of $|\Delta F_{\mathrm{b}}|$ should be significantly smaller than that of $|\Delta F_{\mathrm{c}}|$. More importantly, this means there must exist a coexistent state (PL/AS) appearing within $|\Delta F_{\mathrm{b}}|<|\Delta F|<|\Delta F_{\mathrm{c}}|$. However, outside $|\Delta F_{\mathrm{b}}|<|\Delta F|<|\Delta F_{\mathrm{c}}|$, the existing stable states are only PL and AS, respectively.  Moreover, Eq. (\ref{lowerDF}) shows that the existent criterion for PL/AS states is dependent on the damping constant, i.e. there exists an upper limit of $\alpha$ that ensures $|\Delta F_{\mathrm{b}}|=|\Delta F_{\mathrm{c}}|$, i.e.
\setlength\abovedisplayskip{6pt}
\setlength\belowdisplayskip{6pt}
\begin{eqnarray}
\alpha_{\mathrm{c}}=\frac{2\pi}{S'}\sqrt{\frac{g_{0}}{2}},
\label{criticaldamp}
\end{eqnarray}
below which the existence of the PL/AS state can be assured.

\begin{figure*}
\begin{center}
\includegraphics[width=16cm]{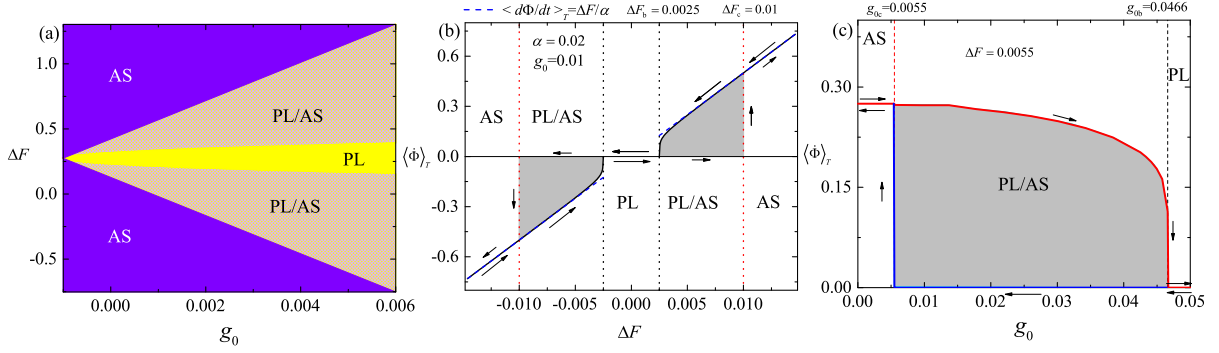}
\end{center}
\caption{(Color online) Stable dynamical states of a phase-locked pendulum. (a)Phase diagram of the stable dynamical states as a function of $g_{0}$ and the driving force $\Delta F$. Here the purple, yellow, and yellow regions with a purple dense pattern denote the AS, PL, and coexisting states (PL/AS), respectively. Also, the damping constant is taken to be $\alpha=0.02$. Note that, $\alpha<\alpha_{\mathrm{c}}=0.0785$.
(b)Hysteretic loop of $\langle\dot{\Phi}\rangle_{T}$ against $\Delta F$, which is highlighted by the gray color. The red and black vertical
 dot lines mark the threshold values of $\Delta F$, i.e. $\Delta F_{\mathrm{b}}=0.0025$ and $\Delta F_{\mathrm{c}}=0.0466$, respectively. The blue dash line denotes the asymptotic line of $\dot{\Phi}$ as a function of $\Delta F$, i.e. $\langle\dot{\Phi}\rangle_{T}=\Delta F/\alpha$.
   (c)Hysteretic loop of $\langle\dot{\Phi}\rangle_{T}$ against $g_{0}$. The red and black
vertical dash lines mark the critical values of $g_{0}$, i.e. $g_{0\mathrm{c}}=0.0055$ and $g_{0\mathrm{b}}=0.0466$, respectively. Here, $\Delta F=0.0055$. }
\label{pendulumhyster}
\end{figure*}

We would like to emphasize here that due to the existence of PL/AS state, the phase-locking dynamics of a pendulum is very different from that of Adler's equation\cite{Adler1973}. The PL/AS state also implies that even if $|\Delta F|<|\Delta F_{\mathrm{c}}|$ is satisfied, there still exist some initial states that eventually evolve into AS states\cite{Bonin2009,Tabor2010,DongLi2011,DAquino2017}. This can be explained physically as follows. For a large enough frequency mismatch $|\Delta\omega|=|(F_{\textrm{dc}}/\alpha)-\omega_{e}|$, i.e. $|\Delta F_{\mathrm{b}}|<|\Delta F|<|\Delta F_{\mathrm{c}}|$, there must exist some initial states where particle gains
sufficient kinetic energy from $\Delta F$ to permanently escape from the trap of the locking potential. As indicated in Fig. \ref{initialstate}, we take two initial states apparently evolving into AS states as examples. One has a higher initial velocity than the terminal one $|\langle\dot{\Phi}\rangle_{T}|$ and is parallel to $\Delta F$; the other is initially static at the point approaching $\Phi_{\mathrm{un}}$ from its right (left) side for positive (negative) $\Delta F$. Here, $\Phi_{\mathrm{un}}$ corresponds to the maximum of the potential $U_{\mathrm{eff}}(\Phi)$. In other words, due to the assistance of sufficient kinetic energy with $\alpha<\alpha_{\mathrm{c}}$ at these states, $\Delta F_{\mathrm{b}}$ can be further lowered compared to $\Delta F_{\mathrm{c}}$. Finally, we would like to point out that due to the relationship between the non-linear frequency shift and kinetic-like energy, for any non-linear auto-oscillators, e.g. STNOs, the presence of the PL/AS state should be their common characteristic.

As shown in Fig. \ref{pendulumhyster}(a), one can analytically obtain the phase diagram of stable dynamic states as a function of $g_{0}$ and $\Delta F$ from Eqs. (\ref{PLDF}), (\ref{lowerDF}), and (\ref{criticaldamp}). Moreover, when $\Delta F$ is shifted through the PL/AS state back and forth along the direction of a constant $g_{0}$ at a very slow pace, a hysteretic frequency mismatch response $\langle\dot{\Phi}\rangle_{T}$ can be observed, as shown by the gray areas of Fig. \ref{pendulumhyster}(b). Here, the black arrows indicate the process of hysteretic loops. When $\Delta F$ increases from PL state, i.e. $|\Delta F|<|\Delta F_{\mathrm{b}}|$ the particle will stay in PL state until $\Delta F=\pm\Delta F_{\mathrm{c}}$. Once $|\Delta F|>|\Delta F_{\mathrm{c}}|$, an abrupt jump from non-zero $\langle\dot{\Phi}\rangle_{T}$ occurs, and the dependence of $\langle\dot{\Phi}\rangle_{T}$ on $\Delta F$ can be linearly approximated by $\langle\dot{\Phi}\rangle_{T}=\Delta F /\alpha$. However, when $\Delta F$ reduces from $|\Delta F|>
|\Delta F_{\mathrm{c}}|$, the particle will stay in the AS state until $|\Delta F|=|\Delta F_{\mathrm{b}}|$, and a non-linear
dependence of $\langle\dot{\Phi}\rangle_{T}$ on $\Delta F$ appears, which is due to the locking potential. By solving Eq. (\ref{criterAS}) and using $T(p)=\int_{0}^{2\pi}d\Phi/\sqrt{2p+2g_{0}\cos(\Phi-\Phi_{0})}$, one can analytically obtain $\langle\dot{\Phi}\rangle_{T}$ as a function of $\Delta F$, as indicated by the black solid curve in Fig. \ref{pendulumhyster}(b). Notably, for the trajectories with an energy close to $g_{0}$,
the particle will spend much more time traversing the maximum point of the potential than its minimum point, resulting in
a much lower $\langle\dot{\Phi}\rangle_{T}$ relative to $\langle\dot{\Phi}\rangle_{T}=\Delta F /\alpha$.

Similarly, when one changes the amplitude of $g_{0}$ along the direction of a constant $\Delta F$ at a very slow pace, a hysteretic response $\langle\dot{\Phi}\rangle_{T}$ can be also observed, as shown in Fig. \ref{pendulumhyster}(c). Also, at the two ends of the hysteretic loop marked by the gray area, there exist two critical values of $g_{0}$. One indicates the bifurcation occurring at the border between AS and PL/AS states ($g_{0}=g_{0\mathrm{c}}$); the other indicates the one occurring at the border between PL/AS and AS states ($g_{0}=g_{0\mathrm{b}}$). In these, $g_{0\mathrm{c}}$ is smaller than $g_{0\mathrm{b}}$, meaning that for the locking potential with a higher energy barrier the more economic way to stimulate the AS state is relying on the assistance of kinetic energy rather than lowering the barrier hight. To summarise, the hysteretic phase-locking criterion confirms that the PL/AS state appears between the PL and AS states.

In addition to the initial state sensitivity of PL states, the pendulum-like equation also predicts the transient regime of PL states, different from that of Adler's equation. By linearizing Eq. (\ref{pendulobygrb}) about the stable phase-locked angle $\Phi_{\mathrm{PL}}$, i.e. $\delta\Phi\equiv\Phi-\Phi_{\mathrm{PL}}$, one can easily obtain
\setlength\abovedisplayskip{6pt}
\setlength\belowdisplayskip{6pt}
\begin{eqnarray}
\ddot{\delta\Phi}+\alpha\dot{\delta\Phi}
+\omega_{0}^{2}\delta\Phi=0.
\label{linear}
\end{eqnarray}
Here, $\omega_{0}=\sqrt{g_{0}}[1-(\Delta F/g_{0})^{2}]^{1/4}$. In the \textit{under} damped case, i.e. $\omega_{0}>\alpha/2$, we obtain an oscillatory decaying solution $\delta\Phi(t)=C_{0}e^{-\alpha t/2}\cos(\omega't+C_{1})$, with $\omega'=\sqrt{4\omega_{0}^{2}-\alpha^{2}}/2$. This means that the transient time of PL states is decided only by $\alpha$ rather than $g_{0}$. In the \textit{critical} damped case, i.e. $\omega_{0}=\alpha/2$, the solution is $\delta\Phi(t)=(C_{0}+C_{1}t)e^{-\alpha t/2}$, where the oscillatory regime starts to disappear. In the \textit{over} damped case, i.e. $\omega_{0}<\alpha/2$, the solution becomes $\delta\Phi(t)=(C_{0}e^{-i\omega't}+C_{1}e^{i\omega't})
e^{-\alpha t/2}$, where the decaying transient time is dependent on $\alpha$, $g_{0}$, as well as the initial states. Notably, in this case, the initial state with more potential energy or less kinetic energy, i.e. $|C_{1}|>|C_{0}|$, will have a longer transient time scale.
\\

Moreover, due to the periodicity of the anisotropic force in $\phi$, Eq. (\ref{pendulobygra}) can be generalized to the more complex case by the Fourier expansion
\setlength\abovedisplayskip{6pt}
\setlength\belowdisplayskip{6pt}
\begin{eqnarray}
\ddot{\phi}+\alpha\dot{\phi}
&=&F_{\mathrm{dc}}-\sum_{n=0}^{\infty}g_{n0}\sin \left(n\phi-\omega_{e}t-\Phi_{n0}\right).
\label{pendulobygra1}
\end{eqnarray}
Here, we would like to point out that the expansion coefficients $g_{n0}$ contains not only the information about the form of the anisotropic force, but also about the geometry of the conserved trajectory. As an example, if the conservative trajectory has no axial symmetry, the projection of the force with even the simplest from, e.g. a uniform gravity, on the trajectory will also take a complicated form.

In the following, we extend the case of an individual phase-locked pendulum to that of an asymmetric pair of mutually phase-locking pendulums, which can be used to analyze the mutual phase-locking of STNO pairs. First of all, the governing
equations for pendulum pairs take the following form:
\setlength\abovedisplayskip{6pt}
\setlength\belowdisplayskip{6pt}
\begin{eqnarray}
\ddot{\phi}_{1}+\alpha_{1}\dot{\phi}_{1}&=&F_{1}-g_{0}\sin \left(\phi_{1}-\phi_{2}\right),\nonumber\\
\ddot{\phi}_{2}+\alpha_{2}\dot{\phi}_{2}&=&F_{2}-g_{0}\sin \left(\phi_{2}-\phi_{1}\right).
\label{pendupair}
\end{eqnarray}
Here, $\alpha_{1}\neq\alpha_{2}$ and $F_{1}\neq F_{2}$. Also, the form of the coupling force means that the two pendulums
become each other's locking potential sources, emitting effective circularly polarized oscillating uniform gravitational forces on the otherl pendulum, respectively. Moreover, using a new set of variables $\phi_{+}\equiv\phi_{1}+\phi_{2}$ and $\phi_{-}\equiv\phi_{1}-\phi_{2}$, Eq. (\ref{pendupair}) becomes
\setlength\abovedisplayskip{6pt}
\setlength\belowdisplayskip{6pt}
\begin{eqnarray}
\ddot{\phi}_{+}+\left(\frac{\alpha_{+}}{2}\right)\dot{\phi}_{+}+\left(\frac{\alpha_{-}}{2}\right)\dot{\phi}_{-}
&=&F_{+},\nonumber\\
\ddot{\phi}_{-}+\left(\frac{\alpha_{+}}{2}\right)\dot{\phi}_{-}+\left(\frac{\alpha_{-}}{2}\right)\dot{\phi}_{+}&=
&F_{-}-g_{0}\sin \phi_{-},
\label{pendupluminu}
\end{eqnarray}
where $\alpha_{\pm}\equiv\alpha_{1}\pm\alpha_{2}$ and $F_{\pm}\equiv F_{1}\pm F_{2}$. According to Eq. (\ref{pendupluminu}), we know that the rotation excitation and phase-locking of the pendulum pairs are governed by
$\phi_{+}$ and $\phi_{-}$ equations, respectively. Accordingly, by comparison with the free-running pendulum equation one can obtain the stable  $\dot{\phi}_{+}$ as follows:
\setlength\abovedisplayskip{6pt}
\setlength\belowdisplayskip{6pt}
\begin{eqnarray}
\dot{\phi}_{+}=\frac{1}{\alpha_{+}}\left(2F_{+}-\alpha_{-}\dot{\phi}_{-}\right).
\label{stablephaseplus}
\end{eqnarray}
By substituting Eq. (\ref{stablephaseplus}) into the $\phi_{-}$ equation defined by Eq. (\ref{pendupluminu}),
one gets the $\phi_{-}$ equation decoupled with $\phi_{+}$:
\setlength\abovedisplayskip{6pt}
\setlength\belowdisplayskip{6pt}
\begin{eqnarray}
\ddot{\phi}_{-}+\alpha'_{+}\dot{\phi}_{-}&=
&F'_{-}-g_{0}\sin \phi_{-},
\label{stablephaseminus}
\end{eqnarray}
Here, $\alpha'_{+}=[\alpha_{+}-(\alpha_{-}^{2}/\alpha_{+})]/2$ and $F'_{-}=F_{-}-\left(\alpha_{-}/\alpha_{+}\right)F_{+}$.
 Comparing Eqs. (\ref{stablephaseminus}) and (\ref{pendulobygrb}), one can easily conclude here that the hysteretic frequency response still exists in mutually phase-locked pendulum pairs. The criteria for PL/AS continue to be
$|F'_{-\mathrm{b}}|<|F'_{-}|<|F'_{-\mathrm{c}}|$ and $\alpha'_{+}<\alpha'_{+\mathrm{c}}$, where we only need to replace $\alpha$ and $\Delta F$ in Eqs. (\ref{PLDF}) and (\ref{lowerDF}) by $\alpha'_{+}$ and $F'_{-}$, respectively.

\subsection{\label{C}Pendulum-like Model for Coupled PERP-STNO Pairs}
\begin{figure*}
\begin{center}
\includegraphics[width=9cm]{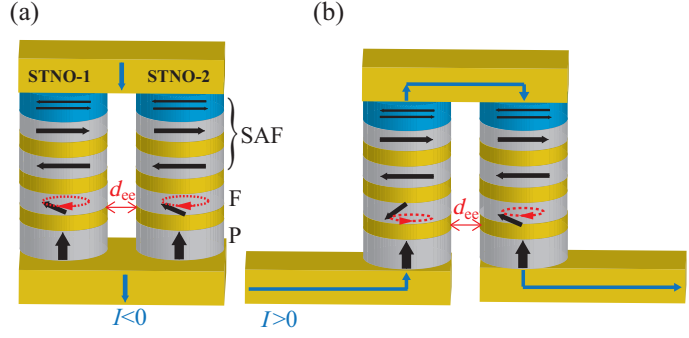}
\end{center}
\caption{(Color online) Schematics of a PERP-STNO pair that are (a) in the parallel and (b) in the serial connections, respectively. Here, P, F, and SAF indicate the pinned, free, and synthetic anti-ferromagnetic layers, respectively.  $I$ is an injected current and $d_{\mathrm{ee}}$ is an edge-to-edge separation. Here, the two red arrows mark the respective precession directions of the two free layer moments.}
\label{Sericonnect}
\end{figure*}
As depicted in Fig. \ref{Sericonnect}, we consider two types of electrically connected nano-pillar-based PERP-STNO pairs with an edge-to-edge separation $d_{\mathrm{ee}}$. One type is connected in parallel (Fig. \ref{Sericonnect}(a)), and the other type is connected in series (Fig. \ref{Sericonnect}(b)). Each pillar is composed of a spin polarizer layer (P) with a perpendicular-to-plane magnetization, a free layer (F) with an in-plane magnetization, and a synthetic antiferromagnetic (SAF) trilayer as an analyzer on top of the free layer.
The thickness of the analyzing layer is designed to be one-tenth of the spin diffusion length\cite{Houssameddine2007}, so the reflecting STT from the analyzer is relatively much smaller than that of the P layer to be neglected. Here we would like to first stress that in the parallel case the same injected current direction implies the pair of free layer magnetic moments precess with the same direction\cite{KJLee2005,Houssameddine2007,JHChang2011,HaoHsuan2017}.
In contrast, in the serial case the two moments will precess in opposite directions due to the opposite injected currents.
\\

Here we assume thatl the magnetization dynamics of the free layers are governed by the macrospin model, i.e. the  Landau-Lifshitz-Gilbert-Slonczewski (LLGS) equation containing the STT effect, as detailed in Refs. \cite{HaoHsuan2016,Chen2019}. Compared to Eq. (\ref{vectformappro}), the vectors $\mathbf{x}$ and $\mathbf{n}$ have been both replaced by the magnetization unit vector $\mathbf{m}$ in the LLGS equation.
In this model, the scaled total energy density $E$ reads as
\setlength\abovedisplayskip{6pt}
\setlength\belowdisplayskip{6pt}
\begin{eqnarray}
E\left(\mathbf{m}\right)&=&E\left(m_{z}, \phi\right)\nonumber\\
&=&\frac{1}{2} \sum_{i=1}^{2} m_{iz}^{2}+\frac{1}{2}\sum_{i=1(i\neq j)}^{2}U_{I}\left(m_{iz},m_{jz},\phi_{i},\phi_{j}\right),\nonumber\\
\label{EnerginLab}
\end{eqnarray}
where the cylindrical coordinate $(m_{iz},\phi_{i})$ has been used to express the energy. The terms on the right-hand side are the demagnetization and interaction energies, respectively. For $U_{I}$, we here take the magnetic dipolar interaction as an example\cite{HaoHsuan2016,HaoHsuan2018}:
\setlength\abovedisplayskip{6pt}
\setlength\belowdisplayskip{6pt}
\begin{eqnarray}
U_{I}\left(m_{1z},m_{2z},\phi_{1},\phi_{2
}\right)&=&-\frac{1}{2}A_{\mathrm{disc}}(d_{\mathrm{ee}})\nonumber\\
&&\times\sqrt{
(1-m_{1z}^{2})(1-m_{2z}^{2})}\nonumber\\
&&\times\bigg\{[3\cos(\phi_{1}+\phi_{2})\nonumber\\
&&+\cos(\phi_{1}-\phi_{2})]
+m_{1z}m_{2z}\bigg\},\nonumber\\
\label{dipolEn}
\end{eqnarray}
where $A_{\mathrm{disc}}(d_{\mathrm{ee}})$ indicates that the dipolar coupling strength is calculated from the quasi-uniformly magnetized circular disc model\cite{HaoHsuan2016}. Note that the spin polarization vector $\mathbf{z}$ coincides with the symmetric axis $\mathbf{p}$ of the demagnetization energy, so the STT here can be directly treated as a negative damping term and the Slonczewski's asymmetirc factor in the STT can be absorbed into the negative damping factor, as indicated in Eq. (\ref{vectformappro}).\\

In our study, the lateral dimension of FL is supposed to be $60\times60$ $\textrm {nm}^{2}$ and the thickness $d=3$ $\textrm{nm}$. The standard material parameters of Permalloy $(\textrm{Ni}_{80}\textrm{Fe}_{20})$ are used for the FL: saturation magnetization $M_{s}=866$ $\textrm{emu}/\textrm{cm}^{3}$, and dimensionless quantities of spin-polarization efficiency $P=0.38$, and $\Lambda=1.8$\cite{XiaoJiang2004}.\\

Based on the axial symmetry of an individual PERP-STNO, including the demagnetization energy as well as the spin polarization vector, the canonical momentum $p_{i}\equiv-m_{iz}$. The stable OP precessional states given by $(\partial H_{Ni,0}/\partial p_{i})_{p_{i0}}=0$ and $(\partial^{2} H_{Ni,0}/\partial p_{i}^{2})_{p_{i0}}>0$ (see Appendix \ref{app:1}) are
\setlength\abovedisplayskip{6pt}
\setlength\belowdisplayskip{6pt}
\begin{eqnarray}
p_{i0}&=&\frac{(\Lambda_{i}^{2}+1)}
{2(\Lambda_{i}^{2}-1)}\nonumber\\
&&\frac{-\sqrt{(\Lambda_{i}^{2}+1)^{2}
-4(\Lambda_{i}^{2}-1)(a_{Ji0}P_{i}\Lambda_{i}^{2}/\alpha)}}
{2(\Lambda_{i}^{2}-1)}.\nonumber\\
\label{equipi0}
\end{eqnarray}
As illustrated in the dynamical state phase diagram (see Ref. \cite{Chen2019}), the regions that allow field and current for $p_{i0}$ are $h_{zi}=0$ and $I_{\mathrm{u}i}<I_{i}<I_{\mathrm{c}i}$, respectively. Here $I_{\mathrm{u}i}$ and $I_{\mathrm{c}i}$ are the critical currents
\setlength\abovedisplayskip{6pt}
\setlength\belowdisplayskip{6pt}
\begin{eqnarray}
I_{\mathrm{u}i}&=&-\bigg(\frac{8\pi eM_{s}^{2}V}{\hbar}\bigg)
\bigg(\frac{2\alpha}{P_{i}}\bigg),
\label{Iu}
\end{eqnarray}
and
\setlength\abovedisplayskip{6pt}
\setlength\belowdisplayskip{6pt}
\begin{eqnarray}
I_{\mathrm{c}i}&=&\bigg(\frac{8\pi eM_{s}^{2}V}{\hbar}\bigg)
\bigg(\frac{2\alpha}{P_{i}\Lambda_{i}^{2}}\bigg),
\label{Ic}
\end{eqnarray}
respectively. The stability $H_{Ni,0}^{(2)}(p_{i0})$, positive damping function $S_{i}(p_{i0})$, negative damping function $a_{i}(p_{i0},\mu_{i})$, nonlinear frequency shift coefficient $H^{(2)}_{Oi}(p_{i0})$, as well as the dipolar interaction energy $U_{I}$ at $p_{i0}$ appearing in Eq. (\ref{nonconpenduL}) are
\setlength\abovedisplayskip{6pt}
\setlength\belowdisplayskip{6pt}
\begin{eqnarray}
H^{(2)}_{Ni,0}(p_{i0})&=&1-\bigg(\frac{a_{Ji0}P_{i}
\Lambda_{i}^{2}}
{\alpha}\bigg)\frac{(\Lambda_{i}^{2}-1)}
{[(\Lambda_{i}^{2}+1)-(\Lambda_{i}^{2}-1)p_{i0}]^{2}},\nonumber\\
\label{stability}
\end{eqnarray}
\setlength\abovedisplayskip{6pt}
\setlength\belowdisplayskip{6pt}
\begin{eqnarray}
S_{i}(p_{i0})&=&\frac{1-p_{i0}^{2}}{1+\alpha^{2}},\nonumber\\
&\approx&1-p_{i0}^{2},
\label{posidampfunc}
\end{eqnarray}
\setlength\abovedisplayskip{6pt}
\setlength\belowdisplayskip{6pt}
\begin{eqnarray}
a_{i}(p_{i0},\mu_{i})&=&\left(\frac{1-p_{i0}^{2}}{1+\alpha^{2}}\right)a_{Ji}(-p_{i0}),\nonumber\\
&\approx&(1-p_{i0}^{2})a_{Ji0}\left[\frac{P_{i}\Lambda_{i}^{2}}{(\Lambda_{i}^{2}+1)+(\Lambda_{i}^{2}-1)(-p_{i0})}\right],
\nonumber\\
\label{negadampfunc}
\end{eqnarray}
\setlength\abovedisplayskip{6pt}
\setlength\belowdisplayskip{6pt}
\begin{eqnarray}
H_{Oi}^{(2)}(p_{i0})=1,
\label{generatedfreq}
\end{eqnarray}
and
\setlength\abovedisplayskip{6pt}
\setlength\belowdisplayskip{6pt}
\begin{eqnarray}
U_{I}(p_{10},p_{20},\phi_{1},\phi_{2})&=&-\frac{1}{2}
A_{\mathrm{disc}}
(d_{\mathrm{ee}})\sqrt{(1-p_{10}^{2})
(1-p_{20}^{2})}\nonumber\\
&&\times\big[3\cos(\phi_{1}+\phi_{2})
+\cos(\phi_{1}-\phi_{2})\big],\nonumber\\
\label{UI}
\end{eqnarray}
respectively. Here, $\mu_{i}$ indicates the spin-polarization efficiencies $(P_{i},\Lambda_{i})$. Notably, in terms of Eq. (\ref{stability}), the stability of OP states
 is mainly provided by the demagnetization energy with a \textit{blue frequency shift} ($H^{(2)}_{Oi}(p_{i0})>0$), and the STT is mainly responsible for shifting the position of $p_{i0}$, which is very different from PMA-STNOs with a \textit{red frequency shift} \cite{Taniguchi_2013}. Thus, PERP-STNO can be driven solely by current without the assistance of an external field (see also Ref.\cite{Chen2019}). Compared with Eqs. (\ref{pendulobygra}) and (\ref{pendupair}), the magnetic dipolar coupling forces appearing in Eq. (\ref{UI}) have an anisotropy as a function of both $\phi_{-}$ and $\phi_{+}$, implying that the dipolar coupling can induce the phase-locking both in the parallel and serial connections. Note, also, that the strength of the dipolar coupling depends not only on the separation $d_{\mathrm{ee}}$ but also $p_{i0}$, and thereby, it is also dependent upon current and damping constant.

 Interestingly, due to the blue frequency shift of a PERP-STNO ($H^{(2)}_{Oi}(p_{i0})>0$), its phase-locked angle will coincide with the minima of the locking potential, which is similar to the OP mode in IP-STNOs \cite{Kang2018}, vortex oscillators \cite{AbreuAraujo2015}, and SHE oscillators \cite{Awad2016,Awad2018}. In contrast, if a certain kind of STNO has a red frequency shift ($H^{(2)}_{Oi}(p_{i0})<0$), its phase-locked angle will coincide with the maxima of the locking potential \cite{Wang2018a}, which is anti-parallel to that of the STNO with a blue frequency shift.

Before analyzing the phase-locking of STNO pairs, we have to further simplify Eq. (\ref{nonconpenduL}). Firstly, since $\alpha_{i} S(p_{i0})\ll1$, where $\alpha_{1,2}$ are both taken as $0.02$, the $2^{\mathrm{nd}}$ term on the right-hand side of Eq. (\ref{nonconpenduL}) can be reasonably neglected. Secondly, from the expression of the last term on the right-hand side of Eq. (\ref{nonconpenduL}) for STNO-1 as an example,
\setlength\abovedisplayskip{6pt}
\setlength\belowdisplayskip{6pt}
\begin{eqnarray}
\sum_{l=1}^{n}\frac{\partial}{\partial\phi_{l}}\left(\frac{\partial U_{I}}{\partial p_{i}}\right)_{p_{i0}}
\dot{\phi}_{l}&=&-\frac{1}{2}[A_{\mathrm{disc}}(d_{\mathrm{ee}})]p_{10}\sqrt{\frac{1-p_{20}^{2}}{1-p_{10}^{2}}}\nonumber\\
&&\times[(3\sin\phi_{+}+\sin\phi_{-})\dot{\phi}_{1}\nonumber\\
&&+(3\sin\phi_{+}-\sin\phi_{-})\dot{\phi}_{2}],\nonumber
\end{eqnarray}
one can easily deduce two points for the parallel connection. First, $\phi_{+}$ increases with time faster than that of $\phi_{-}$, so $\langle\sin\phi_{+}\rangle_{T}\approx0$ in the time order of $\phi_{-}$. Second, for the phase-locking state due to $\dot{\phi}_{1}\approx\dot{\phi}_{2}$, the term associated with $\sin\phi_{-}$ is also reasonably neglected. As for the serial connection, the situation is also similar. Thus,
for the stable phase-locking of STNO pairs, the last term on the right-hand side of Eq. (\ref{nonconpenduL}) can be reasonably neglected. Interestingly, this also reflects that the way that the coupling mechanism affects the STNO frequency mainly relies on shifting the canonical momentum $p_{i}$, rather than the phase angle $\phi_{i}$ directly (see also Appendix \ref{app:2})\cite{Li2010}.

 Subsequently, Eq. (\ref{nonconpenduL}) for the pair of STNOs can be turned into
\setlength\abovedisplayskip{6pt}
\setlength\belowdisplayskip{6pt}
\begin{eqnarray}
\ddot{\phi}_{+}+\left(\frac{\alpha_{\mathrm{eff}+}}{2}\right)\dot{\phi}_{+}+\left(\frac{\alpha_{\mathrm{eff}-}}{2}\right)
\dot{\phi}_{-}&=&F_{\mathrm{eff}+}-g_{0+}\sin\phi_{+},\nonumber\\
\ddot{\phi}_{-}+\left(\frac{\alpha_{\mathrm{eff}+}}{2}\right)\dot{\phi}_{-}+\left(\frac{\alpha_{\mathrm{eff}-}}{2}\right)
\dot{\phi}_{+}&=&F_{\mathrm{eff}-}-g_{0-}\sin \phi_{-},\nonumber\\
\label{pendupairSTNO}
\end{eqnarray}
where $\alpha_{\mathrm{eff}\pm}=\alpha_{\mathrm{eff},1}(p_{10})\pm\alpha_{\mathrm{eff},2}(p_{20})$, $F_{\mathrm{eff}\pm}=F_{\mathrm{eff},1}(p_{10})\pm F_{\mathrm{eff},2}(p_{20})$, $g_{0-}=A_{\mathrm{disc}}(d_{\mathrm{ee}})\sqrt{(1-p_{10}^{2})(1-p_{20}^{2})}$, and $g_{0+}=3A_{\mathrm{disc}}(d_{\mathrm{ee}})\sqrt{(1-p_{10}^{2})(1-p_{20}^{2})}$. In the following, we will analyze the mutual synchronization of a non-identical pair of PERP-STNOs with different sets of spin-polarization efficiencies, i.e. $(\Lambda_{1},P_{1})=(2,0.38)$ and $(\Lambda_{2},P_{2})=(1.8,0.44)$.

By comparison with Eq. (\ref{pendupluminu}), one can find the equations for the OP precession excitation in Eq. (\ref{pendupairSTNO}), e.g. the $\phi_{+}$ equation for the parallel case, and show the existence of threshold driving forces, i.e. threshold currents. That is, near the threshold current, one can reasonably assume $\dot{\phi}_{1}=\dot{\phi}_{2}\approx0$, namely, $\dot{\phi}_{\pm}\approx0$ for the serial $(+)$ and parallel $(-)$ connections, respectively. Then, the equations for trigging precession will be
\setlength\abovedisplayskip{6pt}
\setlength\belowdisplayskip{6pt}
\begin{eqnarray}
\ddot{\phi}_{\pm}+\left(\frac{\alpha_{\mathrm{eff}+}}{2}\right)\dot{\phi}_{\pm}
&\approx&F_{\mathrm{eff}\pm}-g_{0\pm}\sin\phi_{\pm}
\label{paraOPexci}
\end{eqnarray}
for the parallel $(+)$ and serial $(-)$ cases, respectively.

Interestingly, compared to Eq. (\ref{pendulobygrb}), there are two kinds of threshold currents to trigger OP precession:
\setlength\abovedisplayskip{6pt}
\setlength\belowdisplayskip{6pt}
\begin{eqnarray}
|F_{\mathrm{eff\pm,c}}|=g_{0\pm}
\label{thresoldcurrc}
\end{eqnarray}
and
\setlength\abovedisplayskip{6pt}
\setlength\belowdisplayskip{6pt}
\begin{eqnarray}
|F_{\mathrm{eff\pm,b}}|=\left(\frac{\alpha_{\mathrm{eff}+}}{2}\right)\frac{\sqrt{2g_{0\pm}}}{2\pi}S'
\label{thresoldcurrb}
\end{eqnarray}
for the parallel $(+)$ and serial $(-)$ cases, respectively. From these, one can straightforwardly solve for two kinds of threshold currents $|I_{\mathrm{c,p(s)}}|$ and $|I_{\mathrm{b,p(s)}}|$ for the two cases, respectively. Here, p and s appearing in the subscript denote the parallel and serial cases, respectively. Thus, just as the analysis mentioned above, if the criteria are satisfied as follows:
\setlength\abovedisplayskip{6pt}
\setlength\belowdisplayskip{6pt}
\begin{eqnarray}
\alpha_{\mathrm{eff+}}<\alpha_{\mathrm{OP,c}}=\frac{4\pi}{S'}\sqrt{\frac{g_{0\pm}}{2}},
\label{criticaldampOP}
\end{eqnarray}
there exists a coexisting state (static (S)/PL state) in the parallel $(+)$ and serial $(-)$ cases, respectively.
The S/PL state appearing between the S and PL states implies a hysteretic OP precessional frequency response occurs, just as shown in Fig. \ref{pendulumhyster}(b).

Moreover, once the criterion for OP precession is satisfied, $\phi_{\pm}$ will increase (decrease) very fast with time such that $\langle\sin\phi_{\pm}\rangle_{T}\approx0$
in the time order of $\phi_{\mp}$ in the parallel $(-)$ and serial $(+)$ cases, respectively.
Thus, at the stable states $(\ddot{\phi}_{\pm}\approx0)$ we get the stable $\dot\phi_{\pm}$ from Eq. (\ref{pendupairSTNO})
\setlength\abovedisplayskip{6pt}
\setlength\belowdisplayskip{6pt}
\begin{eqnarray}
\dot{\phi}_{\pm}=\frac{1}{\alpha_{\mathrm{eff+}}}\left(2F_{\mathrm{eff}\pm}-\alpha_{\mathrm{eff}-}\dot{\phi}_{\mp}\right),
\label{stablephasepluminu}
\end{eqnarray}
for the parallel $(+)$ and serial $(-)$ cases, respectively. By following the approach mention previously with Eq. (\ref{stablephaseminus}),
one gets the equation governing the phase-locking
\setlength\abovedisplayskip{6pt}
\setlength\belowdisplayskip{6pt}
\begin{eqnarray}
\ddot{\phi}_{\mp}+\alpha'_{\mathrm{eff}+}\dot{\phi}_{\mp}&=
&F'_{\mathrm{eff}\mp}-g_{0_{\mp}}\sin \phi_{\mp}
\label{stablephaselocking}
\end{eqnarray}
for the parallel $(-)$ and serial $(+)$ cases, respectively. Here, $\alpha'_{\mathrm{eff}+}=[\alpha_{\mathrm{eff}+}-(\alpha_{\mathrm{eff}-}^{2}/\alpha_{\mathrm{eff}+})]/2$ and $F'_{\mathrm{eff}\mp}=F_{\mathrm{eff}\mp}-\left(\alpha_{\mathrm{eff}-}/\alpha_{\mathrm{eff}+}\right)
F_{\mathrm{eff}\pm}$. Similarly, one can straightforwardly obtain the critical currents triggering AS state
\setlength\abovedisplayskip{6pt}
\setlength\belowdisplayskip{6pt}
\begin{eqnarray}
|F'_{\mathrm{eff\mp,c}}|=g_{0\mp}
\label{criticacurrc}
\end{eqnarray}
and
\setlength\abovedisplayskip{6pt}
\setlength\belowdisplayskip{6pt}
\begin{eqnarray}
|F'_{\mathrm{eff\mp,b}}|=\frac{\alpha'_{\mathrm{eff}+}\sqrt{2g_{0\mp}}}{2\pi}S'
\label{criticacurrb}
\end{eqnarray}
for the parallel $(-)$ and serial $(+)$ cases, respectively.  From these, two kinds of critical currents $|I'_{\mathrm{c,p(s)}}|$ and $|I'_{\mathrm{b,p(s)}}|$ can be solved for the two cases, respectively. The criteria of the PL/AS state for the parallel $(-)$ and serial $(+)$ are
\setlength\abovedisplayskip{6pt}
\setlength\belowdisplayskip{6pt}
\begin{eqnarray}
\alpha'_{\mathrm{eff+}}<\alpha_{\mathrm{AS,c}}=\frac{2\pi}{S'}\sqrt{\frac{g_{0\mp}}{2}},
\label{criticaldaAS}
\end{eqnarray}
respectively.

As mentioned in section \ref{B}, to calculate the phase-locked frequency of STNO pairs, one can solve the effective energy ($p_{0}>g_{0\pm}$) of the stable PL state from Eq. (\ref{paraOPexci}) as follows:
\setlength\abovedisplayskip{6pt}
\setlength\belowdisplayskip{6pt}
\begin{eqnarray}
\mp F_{\mathrm{eff}\pm}&=&\left(\frac{\alpha_{\mathrm{eff}+}}{2}\right) S_{\pm}(p_{0})
\label{stablepforPL}
\end{eqnarray}
for the parallel $(+)$ and serial $(-)$ cases, respectively. Here, $S_{\pm}(p)=(1/2\pi)\int_{0}^{2\pi}d\phi_{\pm}\sqrt{2p
 +2g_{0\pm}\cos\phi_{\pm}}$ are the damping functions for the two cases, respectively. Subsequently, by substituting solution $p_{0}$ of Eq. (\ref{stablepforPL}) into
\setlength\abovedisplayskip{6pt}
\setlength\belowdisplayskip{6pt}
\begin{eqnarray}
\langle\dot{\phi}_{\pm}\rangle_{T}=\frac{(\pm)(2\pi)}
{\int_{0}^{2\pi}\frac{d\phi_{\pm}}{\sqrt{2p_{0}+2g_{0\pm}\cos\phi_{\pm}}}},
\label{PLangularf}
\end{eqnarray}
one gets the phase angular sum (difference) frequencies of
PL state for the two cases, respectively. Here, when $F_{\mathrm{eff}\pm}$ is negative, a minus sign must be added to the numerator on the right-hand side of Eq. (\ref{PLangularf}). Accordingly, the frequencies of the STNO pairs in the PL state are
\setlength\abovedisplayskip{6pt}
\setlength\belowdisplayskip{6pt}
\begin{eqnarray}
f_{1}=f_{2}=f_{\mathrm{PL}}=\frac{4\pi M_{s}\gamma}{2(2\pi)}|\langle\dot\phi_{\pm}\rangle_{T}|
\label{PLfreqSTNO}
\end{eqnarray}
for the two cases, respectively.

Similarly, to calculate the frequencies
of STNO pairs in AS state, one can solve the effective energy ($p_{0}>g_{0\mp}$) of the stable AS state from Eq. (\ref{stablephaselocking}) as follows:
\setlength\abovedisplayskip{6pt}
\setlength\belowdisplayskip{6pt}
\begin{eqnarray}
\mp F'_{\mathrm{eff}\mp}&=&\alpha'_{\mathrm{eff}+} S_{\mp}(p_{0})
\label{stablepforAS}
\end{eqnarray}
for the parallel $(-)$ and serial $(+)$ cases, respectively. Here, $S_{\mp}(p)=(1/2\pi)\int_{0}^{2\pi}d\phi_{\mp}\sqrt{2p
 +2g_{0\mp}\cos\phi_{\mp}}$ are the damping functions for the two cases, respectively. Subsequently, by substituting solution $p_{0}$ of Eq. (\ref{stablepforAS}) into
\setlength\abovedisplayskip{6pt}
\setlength\belowdisplayskip{6pt}
\begin{eqnarray}
\langle\dot{\phi}_{\mp}\rangle_{T}=\frac{(\pm)(2\pi)}
{\int_{0}^{2\pi}\frac{d\phi_{\mp}}{\sqrt{2p_{0}+2g_{0\mp}\cos\phi_{\mp}}}},
\label{ASangularf}
\end{eqnarray}
one gets the phase angular difference (sum) frequencies of AS states for the two cases, respectively. Note that $\langle\dot{\phi}_{\pm}\rangle_{T}$ for these two cases can be obtained from Eq. (\ref{stablephasepluminu}). Finally, the frequencies of the STNO pairs in the AS state are
\setlength\abovedisplayskip{6pt}
\setlength\belowdisplayskip{6pt}
\begin{eqnarray}
f_{1(2)}&=&\frac{4\pi M_{s}\gamma}{2(2\pi)}|\langle\dot{\phi}_{+}\rangle_{T}\pm\langle\dot{\phi}_{-}\rangle_{T}|,
\label{ASfreqSTNO}
\end{eqnarray}
respectively.

In order to estimate the transient time scale of phase-locking, one can first obtain the stable phase-locked angles from Eq. (\ref{stablephaselocking}):
\setlength\abovedisplayskip{6pt}
\setlength\belowdisplayskip{6pt}
\begin{eqnarray}
\phi_{\mp\mathrm{PL}}=\sin^{-1}\left(\frac{F'_{\mathrm{eff}\mp}}{g_{0\mp}}\right)
\label{criticaldaAS}
\end{eqnarray}
for the parallel $(-)$ and serial $(+)$ cases, respectively. By following Eq. (\ref{linear}), one can also obtain the equations of motion around $\phi_{\mp\mathrm{PL}}$ for the two cases:
\setlength\abovedisplayskip{6pt}
\setlength\belowdisplayskip{6pt}
\begin{eqnarray}
\ddot{\delta\phi_{\mp}}+\alpha'_{\mathrm{eff}+}\dot{\delta\phi_{\mp}}
+\omega_{0\mp}^{2}\delta\phi_{\mp}=0,
\label{linearSTNO}
\end{eqnarray}
respectively. Here $\delta\phi_{\mp}\equiv\phi_{\mp}-\phi_{\mp\mathrm{PL}}$ and $\omega_{0\mp}=\sqrt{g_{0\mp}}[1-( F'_{\mathrm{eff}\mp}/g_{0\mp})^{2}]^{1/4}$. Similarly, in the under-damped case, i.e. $\omega_{0\mp}>\alpha'_{\mathrm{eff+}}/2$, the solution of Eq. (\ref{linearSTNO}) is $\delta\phi_{\mp}(t)=C_{0}e^{-\alpha'_{\mathrm{eff}+}t}\cos(\omega'_{\mp}t+C_{1})$, where $\omega'_{\mp}=\sqrt{4\omega_{0\mp}^{2}-\alpha'^{2}_{\mathrm{eff}+}}/2$. In the critically-damped case, i.e. $\omega_{0\mp}=\alpha'_{\mathrm{eff}+}/2$, the solution is $\delta\phi_{\mp}(t)=(C_{0}+C_{1}t)e^{-\alpha'_{\mathrm{eff}+}t/2}$. In the over-damped case, i.e. $\omega_{0\mp}<\alpha'_{\mathrm{eff+}}/2$, the solution is $\delta\phi_{\mp}(t)=(C_{0}e^{-i\omega'_{\mp}t}+C_{1}e^{-i\omega'_{\mp}t})e^{-\alpha'_{\mathrm{eff}+}t/2}$.

\subsection{\label{D}Mutual Synchronization for Parallel and Serial Connections}
\begin{figure*}
\begin{center}
\includegraphics[width=15.5cm]{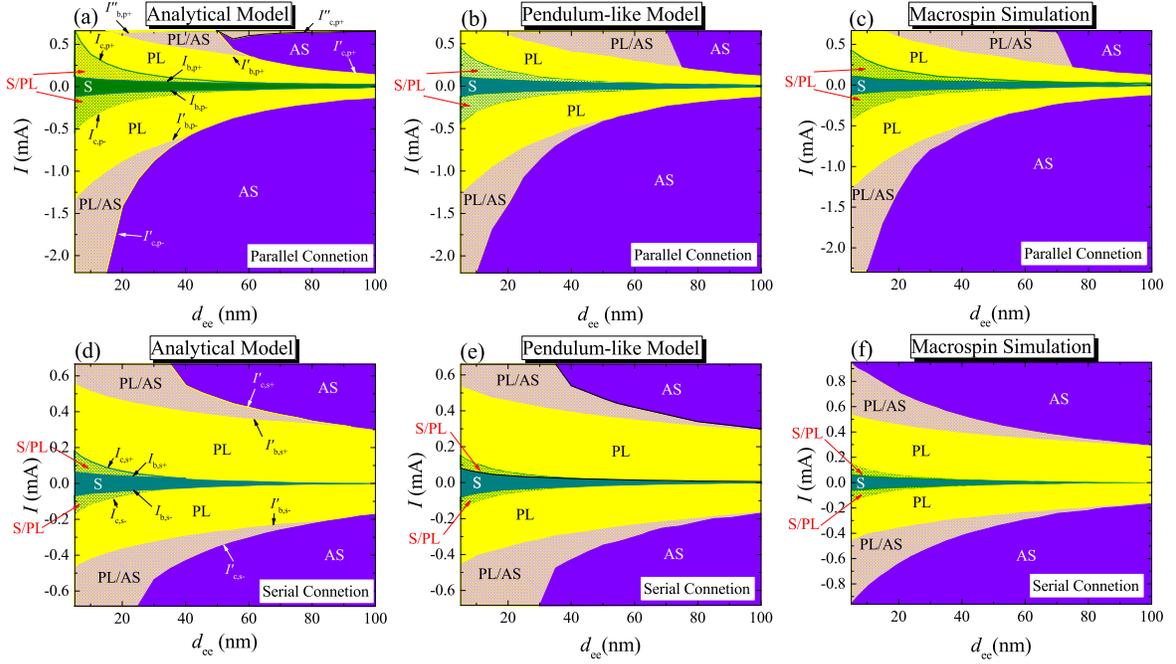}
\end{center}
\caption{(Color online) Phase diagrams for synchronization state as a function of edge-to-edge separation $d_{\mathrm{ee}}$ and current $I$ for the parallel ((a) and (c)) and serial ((b) and (d)) connections, respectively. Diagrams (a) and (b) are calculated from the theoretical model, while (c) and (d) are given by conducting the macrospin simulation. Here, the dark cyan, yellow, and purple regions indicate S, PL, and AS states, respectively. Also, the green and yellow areas with yellow and purple dense patterns indicate S/PL and PL/AS states, respectively.  }
\label{phasdia}
\end{figure*}

\subsubsection{\label{D1}Phase Diagrams of Synchronization State}
\begin{figure*}
\begin{center}
\includegraphics[width=17.5cm]{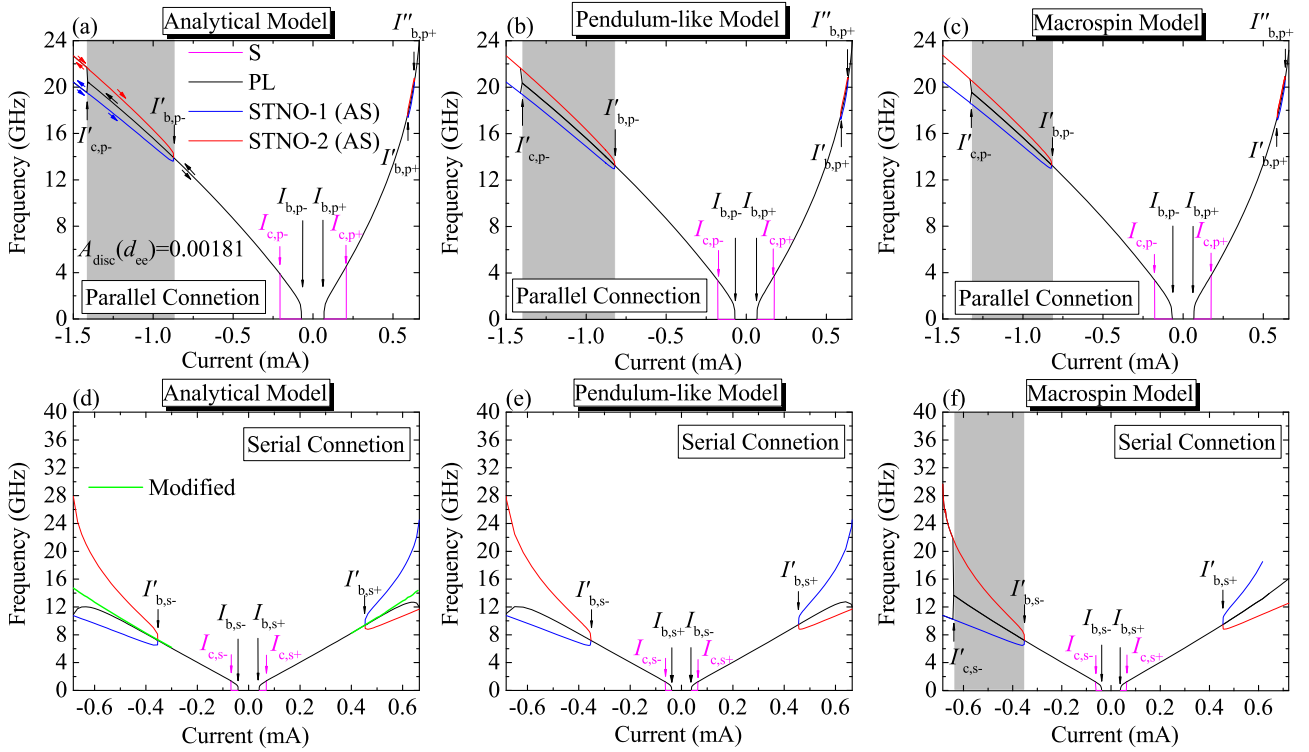}
\end{center}
\caption{(Color online) Hysteric frequency response of phase-locked PERP-STNO pairs against current for the parallel ((a)-(c)) and serial ((d)-(f)) connections, respectively. Here, $d_{\mathrm{ee}}$ and $\alpha$ are taken to be 20 nm and 0.02, respectively. Figures ((a),(d)), ((b),(e)), and ((c),(f)) denote the results of the analytical, pendulum-like, and macrospin models, respectively. The hysteretic phase-locking areas are highlighted by the gray color. The magenta, black, blue, and red curves indicate the frequencies of the S, PL, and AS states for STNO-1,2, respectively. The green curves in (d) indicate the modified data by the macrospin model. The black, red, and blue arrows along the curves in (a) indicate hysteretic process. $I_{\mathrm{b,p(s)\pm}}$ and $I_{\mathrm{c,p(s)\pm}}$ indicate the threshold currents of driving PL state for the parallel and serial connections, respectively. $I'_{\mathrm{b,p(s)\pm}}$, $I''_{\mathrm{b,p+}}$, and $I'_{\mathrm{c,p(s)\pm}}$ are the critical currents of stimulating AS state for these two connections, respectively. The values of these currents can be seen in TABLE \ref{tabelp} and TABLE \ref{tabels}.  }
\label{Hysterfre}
\end{figure*}

\begin{figure*}
\begin{center}
\includegraphics[width=12cm]{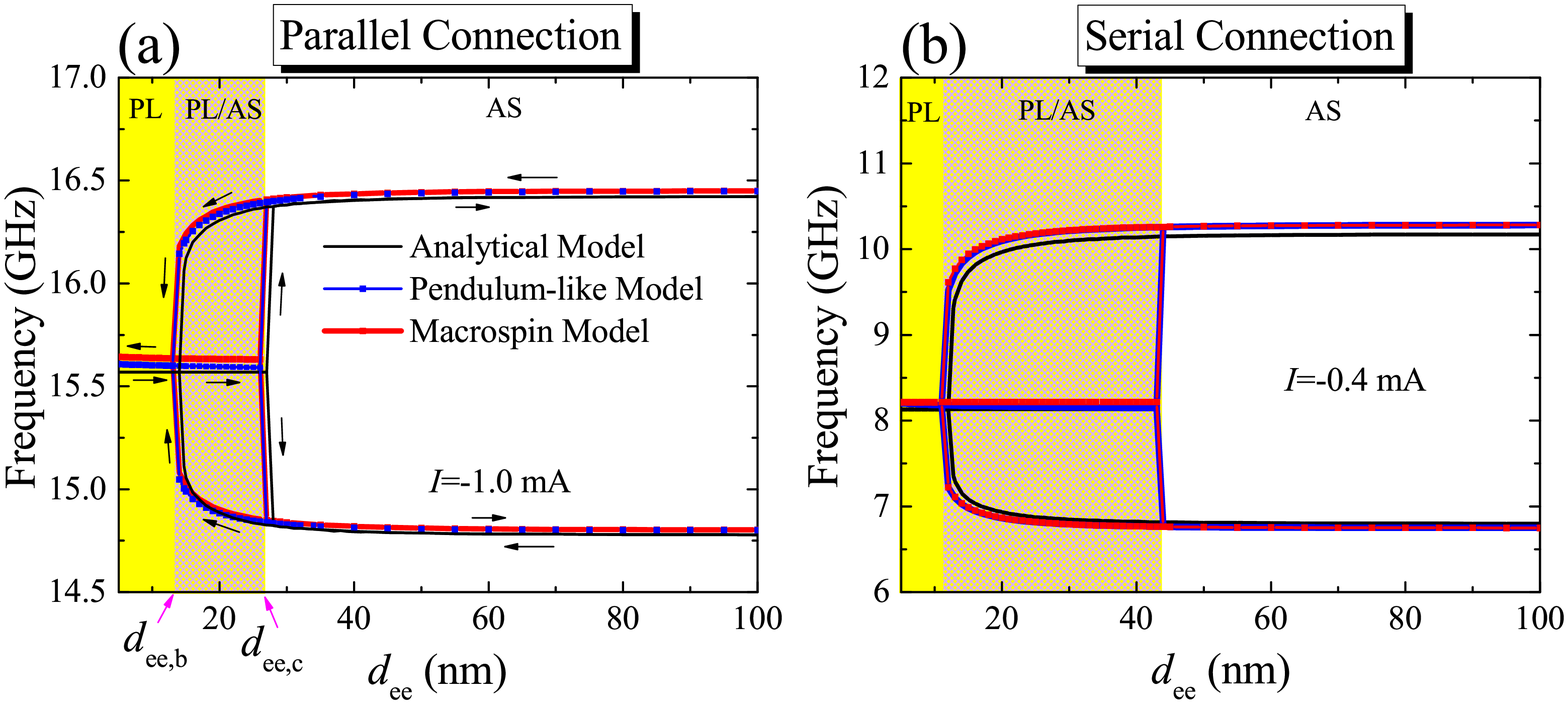}
\end{center}
\caption{(Color online) Hysteretic frequency response of phase-locked PERP-STNO pairs as a function of $d_{\mathrm{ee}}$. Figures (a) and (b) indicate the cases for the parallel and serial connections, respectively. The blue and red square lines denote the theoretical and macrospin simulation results, respectively. The black arrows along the curves indicate the hysteretic process. The gray regions mark the synchronization $d_{\mathrm{ee}}$ regions of the simulation result. In the parallel and serial cases, $I$ are taken to be $-1.0$ mA and $-0.4$ mA, respectively. $d_{\mathrm{ee,b}}$ and $d_{\mathrm{ee,c}}$ are critical separations. In the parallel case, $d_{\mathrm{ee,b}}=14.71$ nm and $d_{\mathrm{ee,c}}=27.2$ nm. In the serial case, $d_{\mathrm{ee,b}}=11.22$ nm and $d_{\mathrm{ee,c}}=43.8$ nm.}
\label{fpvsdee}
\end{figure*}

Using Eqs. (\ref{thresoldcurrc}) to (\ref{criticaldaAS}), one can straightforwardly analytically solve the dynamic phase diagrams of synchronization state as a function of current $I$ and separation $d_{\mathrm{ee}}$ for the parallel and serial connections, respectively. As indicated in Figs. \ref{phasdia}(a) for the parallel case, there exist five types of stable states on the phase plane, namely S, S/PL, PL, PL/AS, and AS states, respectively. These are divided by the threshold and critical currents $I_{\mathrm{b,p}\pm}$, $I_{\mathrm{c,p}\pm}$, $I'_{\mathrm{b,p}\pm}$, $I'_{\mathrm{c,p}\pm}$, $I''_{\mathrm{b,p}+}$, $I''_{\mathrm{c,p}+}$, respectively. Note, here, that according to Eqs. (\ref{Iu}) and (\ref{Ic}) for an individual PERP-STNO, only the current ranging from $-2.2$ mA to $0.66$ mA on the plane can ensure that the STNO pairs are both stimulated.

To verify the analytical result, we supply results for the pendulum-like and the macrospin models for comparison purposes by numerically solving Eq. (\ref{nonconpenduL}) and the LLGS equation. The results are shown in Figs. \ref{phasdia}(b) and (c), respectively. Notice that the analytical result is basically in good agreement with these numerical results, both qualitatively and quantitatively. However, one can still find three differences between the analytical model and the other models. First, a relatively lager AS state area (purple color) in the analytical model for positive current; second, there is a relatively less obvious PL state area (yellow color with a purple dense pattern) appears on top of the PL/AS state for positive current in the two other models; third, the PL/AS state area appearing on top of the AS state for positive current in the analytical model is not seen in the two other models.

In the serial case, as indicated by Fig. \ref{phasdia}(d) the situation is similar. The whole phase plane is divided by the threshold and critical currents $I_{\mathrm{b,s}\pm}$, $I_{\mathrm{c,s}\pm}$, $I'_{\mathrm{b,s}\pm}$, $I'_{\mathrm{c,s}\pm}$, respectively. However, due to opposite current injections, only current ranges from $-0.68$ mA to $0.67$ mA can ensure that the STNO pairs be both trigged, which is the same as the pendulum-like model (Fig. \ref{phasdia}(e)) instead of the macrospin one (Fig. \ref{phasdia}(f)). Thus, in the analytical and pendulum-like models, we fails to predict about the stable dynamics in the current range of $|I|>0.67$ mA.

Notice that in the serial case the phase diagram displays a better symmetry against current than in the parallel one. The reason can be given as follows. We know that in the serial case the frequency mismatch between the STNO pairs is mainly due to the asymmetry of the STT $a_{Ji}(-p_{i0})$ on current direction instead of their inconsistency in spin-polarization efficiency $(P_{i},\Lambda_{i})$. Thus, the frequency mismatch in the positive current is close to that in the negative current. In contrast, in the parallel case the asymmetry of the phase diagram against current is obviously due to the asymmetry of the STT on current direction, and the frequency mismatch is due to the inconsistency in spin-polarization efficiencies between the STNO pairs. Additionally, the values of the threshold currents ($I_{\mathrm{b,s}\pm}$ and $I_{\mathrm{c,s}\pm}$) in the serial case are significantly smaller than those in the parallel one. This reflects that the potential barrier obstructing the trigging of PL states in the parallel case is higher than that in the serial one, as can be seen in Eq. (\ref{pendupairSTNO}).

\subsubsection{Hysteretic Synchronization Frequency Response}
Just as pointed out by Figs. \ref{pendulumhyster}(a) and (b) pointed out, the phase diagrams shown in Fig. \ref{phasdia} has implied the existence of a hysteretic synchronization frequency response, which are given here by the analytical (Eqs. (\ref{PLfreqSTNO}) and (\ref{ASfreqSTNO})), pendulum-like, and macrospin models, respectively, as shown in Fig. \ref{Hysterfre}. For the S/PL states, there are two kinds of frequency responses. The S state has $f_{1,2}=0$ GHz until $I=|I_{\mathrm{b,p(s)}\pm}|$ (see the magenta curves in Fig. \ref{Hysterfre}); the PL state has nonlinear $f_{1,2}$ on $I$ within $|I_{\mathrm{b,p(s)}\pm}|<|I|<|I_{\mathrm{c,p(s)}\pm}|$ (see the black curves), which is similar to the case shown in Fig. \ref{pendulumhyster} (b). For the PL/AS states, one can easily find that there are three response curves that are coexistent. One indicates the PL state with $f_{1}=f_{2}$; the other indicates the AS state with $f_{1}\neq f_{2}$ (see the blue and red curves).

  In the parallel case, when $d_{\mathrm{ee}}$ is taken as $20$ nm, there exist three hysteretic loops, which are surrounded by the threshold currents  $I_{\mathrm{b,p}\pm}$, $I_{\mathrm{c,p}\pm}$ and the critical currents $I'_{\mathrm{b,p}-}$, $I'_{\mathrm{c,p}-}$, respectively, as can be seen in Figs. \ref{Hysterfre}(a)-(c). Here, the values of these currents for the three models are shown in TABLE \ref{tabelp}, indicating that the analytical results are in good agrement with those of the other two models. Note that, except for the loops for trigging PL states as predicted in Eqs. (\ref{thresoldcurrc})-(\ref{criticaldampOP}), only a loop of phase-locking (see the gray areas in Figs. \ref{Hysterfre}(a)-(c)) exists in the negative current, indicating the existence of the PL/AS state. However, this does not mean that the PL/AS state would not appear in the positive current. Just as the analysis in section \ref{B} pointed out, since the PL/AS state in the positive current is not surrounded by the PL and AS states, then the hysteretic loop will not be existent. Based on this, it is not enough to identify all of the PL/AS states by only confirming the presence of hysteretic loops.

\begin{table}
\caption{Threshold and critical currents in parallel case}
\begin{tabular*}{8cm}{llll}
\hline
 (mA)& Analytical & Pendulum-like & Macrospin  \\
\hline
$I_{\mathrm{b,p\pm}}$ &(-0.068,0.068)&(-0.072,0.065)&(-0.068,0.069)\\
$I_{\mathrm{c,p\pm}}$ &(-0.21,0.21)&(-0.18,0.17)&(-0.18,0.17)\\
$I'_{\mathrm{b,p\pm}}$&(-0.87,0.59)&(-0.82,0.59)&(-0.82,0.58) \\
$I''_{\mathrm{b,p+}}$& 0.63 & 0.635 & 0.635 \\
$I'_{\mathrm{c,p-}}$&-1.40&-1.40&-1.32\\
\hline
\label{tabelp}
\end{tabular*}
\end{table}

\begin{table}
\caption{Threshold and critical currents in serial case}
\begin{tabular*}{8cm}{llll}
\hline
  (mA)& Analytical & Pendulum-like & Macrospin  \\
\hline
$I_{\mathrm{b,s\pm}}$&(-0.04,0.04)&(-0.036,0.04)&(-0.037,0.04)\\
$I_{\mathrm{c,s\pm}}$&(-0.068,0.07)&(-0.062,0.064)&(-0.061,0.065)\\
$I'_{\mathrm{b,s\pm}}$&(-0.35,0.453)&(-0.35,0.46)&(-0.35,0.45) \\
$I'_{\mathrm{c,s-}}$&Non&Non&-0.64\\
\hline
\label{tabels}
\end{tabular*}
\end{table}

In the serial case with the same $d_{\mathrm{ee}}$, the situation is also similar, as shown in Figs. \ref{Hysterfre}(d)-(f). Compared to the parallel case, the main differences are that the serial case has a more symmetric frequency response for current, smaller threshold currents $|I_{\mathrm{b,s}\pm}|$ and $|I_{\mathrm{c,s}\pm}|$, much larger frequency mismatches in the AS states, and lower $f_{\mathrm{PL}}$ than average frequency ($f_{\mathrm{av}}=(f_{1}+f_{2})/2$). The reason for the lower $f_{\mathrm{PL}}$ can be easily seen from Eq. (\ref{stablephasepluminu}) as follows: \setlength\abovedisplayskip{6pt}
\setlength\belowdisplayskip{6pt}
\begin{eqnarray}
|\langle\dot{\phi}_{-}\rangle_{T}|&=&2|\langle\dot{\phi}_{-}\rangle_{T,\mathrm{av}}|,\nonumber\\
&=&\left|\frac{2F_{\mathrm{eff},-}}{\alpha_{\mathrm{eff},+}}-\left(
\frac{\alpha_{\mathrm{eff},-}}{\alpha_{\mathrm{eff}+}}\right)\langle\dot{\phi}_{+}\rangle_{T}\right|,\nonumber\\
&\approx&\left|2\langle\dot{\phi}_{-}\rangle_{T,\mathrm{PL}}-\left(\frac{\alpha_{\mathrm{eff},-}}{\alpha_{\mathrm{eff}+}}\right)
\langle\dot{\phi}_{+}\rangle_{T}\right|.\nonumber
\end{eqnarray}
Here $\langle\dot{\phi}_{-}\rangle_{T,\mathrm{av}}$ and $\langle\dot{\phi}_{-}\rangle_{T,\mathrm{PL}}$ denote the average and phase-locked $\langle\dot{\phi}_{-}\rangle_{T}$, respectively.  Since the STNO with a positive angular velocity $(\dot{\phi}_{i}>0)$ has a higher frequency than that of another one, $\langle\dot{\phi}_{+}\rangle_{T}$ is positive both for the positive and negative currents. As for $F_{\mathrm{eff},-}$ and $\alpha_{\mathrm{eff},-}$, they have opposite signs to each other whether in the positive or negative currents, making $f_{\mathrm{PL}}<f_{\mathrm{av}}$. However, in the parallel case $\langle\dot{\phi}_{-}\rangle_{T}$ is close to zero, leading to $f_{\mathrm{PL}}\approx f_{\mathrm{av}}$.

Notably, the hysteretic loop appearing in the negative current can only be seen in the macrospin simulation, as shown in Fig. \ref{Hysterfre} (f). The reason for this has been mentioned section \ref{D1}, namely, the theoretical and pendulum-like models can only deal with the case for when both of the SNTO pairs are triggered. Besides, at the right end of the blue curve (see Fig. \ref{Hysterfre}(f)), since $m_{1z}=-1$ at $I=0.643$ mA, the magnetization of STNO-1 has been stopped by the STT.
However, even when the current exceeds this value, STNO-2 can still be trigged, as shown by the red curve of Fig. \ref{Hysterfre}(f). So, the PL state can also still be stimulated when the current exceeds this value.

 Note that, in the PL state when the current $|I|$ approaches $0.6$ mA, the increasing rate of frequencies $f_{\mathrm{PL}}$ with $I$ exists turning points, as shown in Figs. \ref{Hysterfre}(d) and (e). This inconsistency between the analytical (pendulum-like) and macrospin models is due to the non-linear
 feature of PERP-STNOs, namely, the dependence of (angular) frequency on momentum (amplitude)  ($m_{iz}=-p_{i}=\dot{\phi}_{i}$). In other words, for a coupled pair of non-linear oscillators, the locking of (angular) frequency naturally means the locking of momentum if the dynamic state energy of an individual oscillator is dominant. However, out of the convenience of analysis, the analytical (pendulum-like) models derived here are based on individual STNOs, where the equilibrium points $p_{i0}$ are assumed to be static and not vary with $\dot{\phi}_{i}$. Thus, in the serial case, the difference between $p_{i0}$ and actual locked $p_{i}$ will be further enlarged at the large $I$ due to a significantly larger frequency mismatch than in the parallel case. To illustrate this point, here we replace $p_{i0}$ with the locked $p_{i}=-m_{iz}$ solved by the macrospin simulation.
 Then, one obtains the modified $f_{\mathrm{PL}}$ as a function of $I$ (see the green curve in Fig. \ref{Hysterfre}(d)), which is in good agreement with that of the macrospin model.
 \begin{figure*}
\begin{center}
\includegraphics[width=13.5cm]{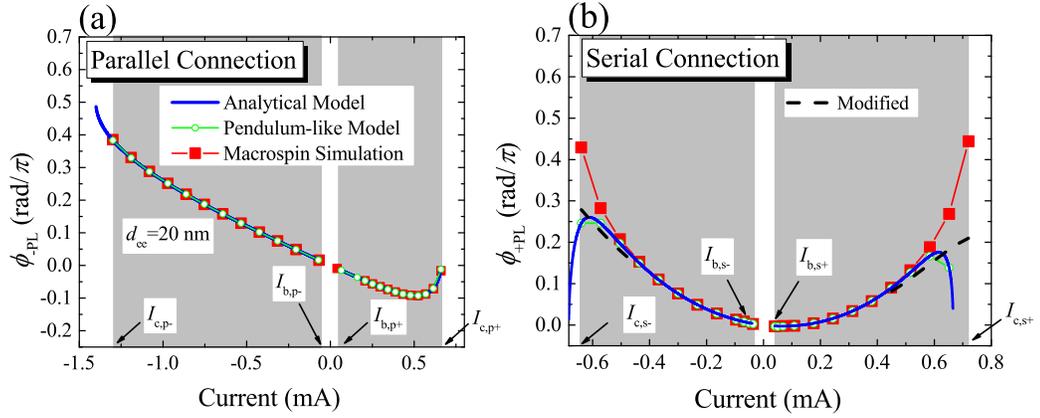}
\end{center}
\caption{(Color online) Stable phase-locked angles $\phi_{\pm\mathrm{PL}}$ as a function of current $I$ for the parallel (a) and serial (b) cases, respectively. Here, $d_{\mathrm{ee}}$ is taken to be 20 nm. The blue solid, green circle, and red square curves indicate the results of the analytical, pendulum-like, and macrospin models, respectively. The black dash line in (b) denotes the modified data of the analytical model by the macrospin simulation. The gray regions mark the PL state of the macrospin simulation result. The values of $I_{\mathrm{b,p(s)}\pm}$ and  $I_{\mathrm{c,p(s)}\pm}$ can be seen in TABLE \ref{tabelp} and TABLE \ref{tabels}, respectively.}
\label{Phasesum}
\end{figure*}

In the following, we also give the hysteretic frequency responses as a function of $d_{\mathrm{ee}}$, which are calculated by the analytical, pendulum-like, and macrospin models, respectively, as shown in Fig. \ref{fpvsdee}. It is obvious that these three results are very consistent with each other. Here, since $A_{\mathrm{disc}}(d_{\mathrm{ee}})$ decreases with increasing $d_{\mathrm{ee}}$, the bifurcation between the PL and PL/AS states occurs at a smaller critical separation $d_{\mathrm{ee,b}}$, which is similar to the case of Fig. \ref{pendulumhyster}(c). Conversely, the bifurcation between the PL/AS and AS states occurs at a larger critical separation $d_{\mathrm{ee,c}}$.
Comparing Figs. \ref{fpvsdee}(a) and (b), one can find that in the serial case the coupling strength is significantly larger than in the parallel case, as can be seen by a significantly larger frequency mismatch as well as $d_{\mathrm{ee},c}=43.8$ nm in the serial case. However, notably, $d_{\mathrm{ee,b}}=14.71$ nm in the parallel case which is larger than $d_{\mathrm{ee,b}}=11.22$ nm in the serial case. This means that
the gap between $d_{\mathrm{ee,b}}$ and $d_{\mathrm{ee,c}}$ is greatly increased in the serial case.  This is due to a much larger $|F'_{\mathrm{eff},+\mathrm{b}}|$ than in the parallel case.

\subsubsection{Phase-locked Angle}
In order to optimize the output power of STNO pairs by phase-locking, it is vital to analyze the stable phase-locked angles $\phi_{\mp\mathrm{PL}}$ for the parallel and serial connections, respectively. As shown in Fig. \ref{Phasesum}, we display $\phi_{\mp\mathrm{PL}}$ as a function of current, which are given by the analytical, pendulum-like, and macrospin models, respectively. Note that, in the simulations the initial states of the two free layer moments has been set as being along the conjunction line between the STNO pairs, which is exactly the same as the stable states formed by the dipolar coupling without driving current. One finds that in the parallel case the analytical and pendulum-like results are verified by the macrospin model very well. However, in the serial case when $|I|$ approaches $0.6$ mA, the increasing rate of $\phi_{+\mathrm{PL}}$ with $I$ appears turning points, inducing the inconsistency between the analytical (pendulum-like) and macrospin models. As discussed previously, the reason for this is that the dependence of (angular) frequency on amplitude (momentum) ($m_{iz}=-p_{i}=\dot{\phi}_{i}$) of PERP-STNOs has not been taken into consideration in the analytical (pendulum-like) model. In order to prove this point in a similar manner to Fig. \ref{Hysterfre}(d), here we obtained the modified result of the analytical model by replacing $p_{i0}$ with the locked $p_{i}=-m_{iz}$ solved by the macrospin simulation, as shown by the black dashed curve of Fig. \ref{Phasesum}(b). In the modified result, notice that the turning points of $\phi_{+\mathrm{PL}}$ have been eliminated.

In addition, due to the inconsistency of spin-polarization efficiencies $(P_{i},\Lambda_{i})$ between the PERP-STNO pairs in thel parallel connection, $\phi_{\mathrm{PL}-}$ displays an odd function of current, as indicated in Fig. \ref{Phasesum}(a). Here we have $a_{1}(p_{10},\mu_{1})<a_{2}(p_{20},\mu_{2})<0$ ($F'_{\mathrm{eff}-}<0$) in the positive current and $a_{1}(p_{10},\mu_{1})>a_{2}(p_{10},\mu_{2})>0$ ($F'_{\mathrm{eff}-}>0$) in the negative current. For the serial connection, due to the asymmetry of the STT strength $a_{Ji}(-p_{i0})$ on current direction, $\phi_{+\mathrm{PL}}$ displays an even function of current, as indicated by Fig. \ref{Phasesum}(b). There we have
 $a_{1}(p_{10},\mu_{1})>0>a_{2}(p_{20},\mu_{2})$ and $|a_{1}(p_{10},\mu_{1})|>|a_{2}(p_{20},\mu_{2})|$ ($F'_{\mathrm{eff}+}>0$) in the positive current and $a_{1}(p_{10},\mu_{1})<0<a_{2}(p_{10},\mu_{2})$ and $|a_{1}(p_{10},\mu_{1})|<|a_{2}(p_{20},\mu_{2})|$ ($F'_{\mathrm{eff}+}>0$) in the negative current.

Moreover, we would like to stress here that the enhancement of the output power of multiple PERP-STNOs is not only dependent on the phase-locked angle $\phi_{\mp\mathrm{PL}}$ but also on the arrangement of analyzers. In this sense, one can straightforwardly deduce from Fig. \ref{Phasesum}(a) that in the parallel connection no matter how the STNO pairs arrange on a plane, as long as their analyzers are parallel to each other the projections of free layer moments on analyzers must be equal to each other for $\phi_{-\mathrm{PL}}\approx0$. However, for the serial connection (see Fig. \ref{Phasesum}(b)), only arranging both of the analyzers along the connection line between the two STNOs can ensure that the equal projections be produced when $\phi_{+\mathrm{PL}}\approx0$. So, it is  concluded here that the parallel connection of PERP-STNO pairs is more easily generalized to multiple PERP-STNOs in a two-dimensional arrangement than the parallel one.

\begin{figure*}
\begin{center}
\includegraphics[width=15cm]{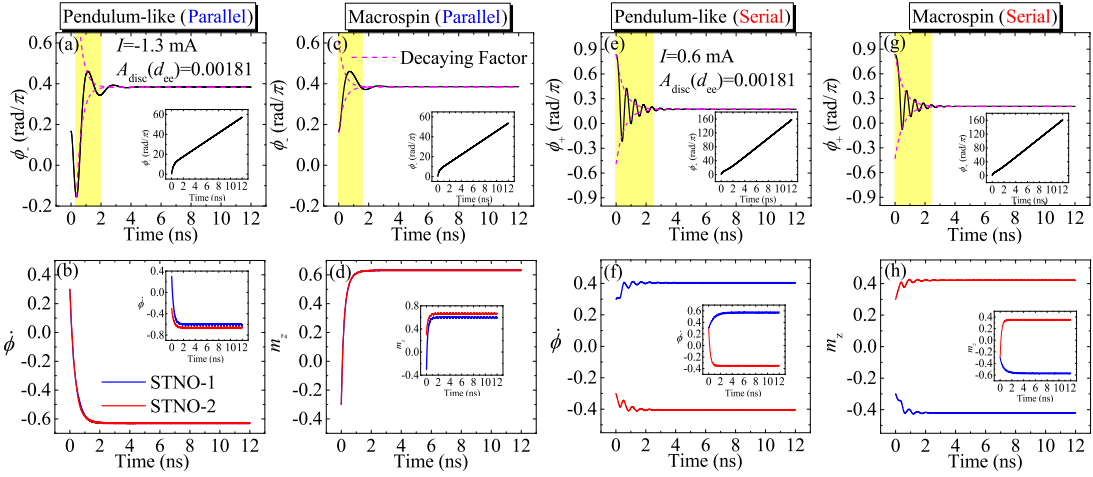}
\end{center}
\caption{(Color online) Transient evolutional processes of mutual synchronization for the parallel ((a)-(d)) and serial ((e)-(h)) connections, respectively, where figures (a), (b), (e), and (f) are the simulations of the pendulum-like model. Figures (c), (d), (g), and (h) are the macrospin simulation results. The upper figures present the transient states for the locking of the phase angle $\phi_{\mp}$. The magenta dashed curves indicate the time trace of the decaying factor $e^{-\alpha_{\mathrm{eff}+}t}$. The transient regimes are highlighted by light yellow colors. The lower figures (b) and (f) show the transient states for the locking of $\dot{\phi}_{1,2}$, and (d) and (h) show the transient states for the locking of $m_{z}$. The panels indicate the time evolution of AS states. The blue and red curves appearing in the lower figures indicate the transient states of STNO-1 and STNO-2, respectively. Here, for the parallel and serial cases, $I$ are taken to be $-1.3$ mA and $0.4$ mA, respectively. $A_{\mathrm{disc}}(d_{\mathrm{ee}})$ is taken to be $0.00181$.}
\label{trainsienstate}
\end{figure*}

\subsubsection{Transient Regime of Synchronization States}
We present the transient states of synchronization in the parallel and serial connections, respectively, in Fig. \ref{trainsienstate}. The numerical results are given by conducting the pendulum-like and macrospin simulations, respectively. As indicated in Fig. \ref{trainsienstate}, the oscillatory transient state regimes for the two kinds of connections both belong to the under damped situation, where the transient time scale of synchronization is around $1-4$ ns and independent on the coupling strength. This has been confirmed both by the pendulum-like and macrospin models. According to Eq. (\ref{linearSTNO}), one can theoretically well-illustrate the condition for the under damped case, namely, $\omega_{0-}=0.0197>\alpha'_{\mathrm{eff}+}/2=0.0076$ for the parallel connection, and $\omega_{0+}=0.0598>\alpha'_{\mathrm{eff}+}/2=0.0049$ for the serial connection. Furthermore, the transient time scale $t_{r}$ of phase-locking can be easily estimated by requiring $e^{-\alpha'_{\mathrm{eff}+}t_{r}}\sim 0.01$, namely, $t_{r}=1.6$ ns for the parallel connection, and $t_{r}=2.453$ ns for the serial connection. Furthermore, the frequencies of oscillatory decaying solutions $f'_{\mp}=\omega'_{\mp}/(2\pi)$ are $f'_{-}=0.555$ $\mathrm{ns}^{-1}$ for the parallel connection, and $f'_{+}=1.817$ $\mathrm{ns}^{-1}$ for the serial connection.

Similar phenomena are also observable in the phase-locking of other types of STNOs\cite{Zhou2010,AbreuAraujo2015,Wang2017}. Notably, we have emphasized previously that for non-linear auto-oscillators the phase-locking of the phase angles also means the locking of the angular velocities $\dot{\phi}_{i}$ (Figs. \ref{trainsienstate}(b) and (f)) or conjugate momentum $m_{iz}$ (Figs. \ref{trainsienstate}(d) and (h)). In the parallel $(+)$ and serial $(-)$ cases, the lockings of $\dot{\phi}_{i}$ and $m_{iz}$ meet with $\dot{\phi}_{1}=\pm\dot{\phi}_{2}$ and $m_{1z}=\pm m_{2z}$, respectively. Subsequently, since due to $\dot{\phi}_{i}=p_{i}=-m_{iz}$ the values of locked angular velocities calculated by the pendulum-like model are very close to those of the macrospin simulation, which can be seen from the locked $m_{iz}$. Additionally, this is also why $m_{iz}$ or amplitude has a corresponding oscillatory decay in transient states, which can be more easily observed in Fig. \ref{trainsienstate}(h).
 Finally, just as emphasized previously, due to the presence of the PL/AS state the synchronization states are actually sensitive to initial states, which can be observed in either pendulum-like or mocrospin simulations (panels in Fig. \ref{trainsienstate}).

\section{\label{sec:3}Summary and Discussion}
In this work, a generalized pendulum-like model is developed based on the two common and fundamental characteristics possessed by non-linear auto-oscillators of all kinds: one is the stability and the other is the non-linear dynamic state energy. Thereby, this new model can be used to elucidate the mechanisms of IC-dependent mutual synchronization of all kinds of non-linear auto-oscillators. Subsequently, we adopt this model to fully analyze the mutual synchronization of PERP-STNO pairs connected in parallel and series, including IC-dependent OP precessional excitation/synchronization as well as the oscillatory decaying regime. These phenomena are all actually induced from the non-linear dynamic state energy, i.e. kinetic-like energy, which cannot be explained by the Kuramoto model. All of the theoretical results are well-verified by the results of numerical simulations.. Furthermore, owing to the blue frequency shift of PERP-STNOs, our synchronization scheme of PERP-STNO pairs can be trigged without the assistance of an external field whether connected in serial or in parallel, which is an advantage compared with other schemes in practical applications.

In the context of application, the presence of IC-dependent mutual synchronization would induce uncertainty to synchronization (see Fig. \ref{phasdia}) and therefore hinder the enhancement of STNOs output power, which should be avoided as much as possible. Fortunately, since the dipolar coupling can control the initial alignments of the free layer moments in PERP-STNOs, the moments would not easily acquire a sufficient amount of kinetic-like energy from the frequency difference between any pair of PERP-STNOs to evolve into AS states after they are driven by current. For example, both the initial stable alignments of the moment pairs must be parallel to the conjunction line between the STNO pairs, making them evolve into the PL state after the current is turned on (see Fig. \ref{Phasesum}). Therefore, we believe our synchronization scheme can effectively overcome this difficulty and therefore be easily extended to an array of multiple oscillators.

\begin{acknowledgments}
The authors gratefully acknowledge the National Natural Science Foundation of China (Grants No. 61627813 and No. 61571023), the International Collaboration Project No. B16001, the National Key Technology Program of China No. 2017ZX01032101, Beihang Hefei Innovation Research Institute Projects (BHKX-19-01) and (BHKX-19-02) for their financial support of this work.
\end{acknowledgments}

\appendix

\section{\label{appa}General Pendulum-like Model for Non-linear Auto-Oscillatory Systems}
\subsection{\label{app:1}Generalized Canonical Cyclic Coordinate}
In general cases, the conserved trajectories of auto-oscillatory systems governed by Eq. (\ref{vectformappro}) do not necessary have a circular shape, which makes the form of $E_{N}$ more complicated for analysis, e.g. IP-STNO\cite{kiselev2003microwave}. That is, in terms of any type of curvilinear coordinate systems $(x,y)$, the conserved part of Eq. (\ref{vectformappro}) generally takes the form\cite{GBertotti2009nonlinear}
\setlength\abovedisplayskip{6pt}
\setlength\belowdisplayskip{6pt}
\begin{eqnarray}
\dot{x}&=&\beta(\mathbf{x})\frac{\partial E_{0}}{\partial y},\nonumber\\
\dot{y}&=&-\beta(\mathbf{x})\frac{\partial E_{0}}{\partial x}.
\label{equaforx}
\end{eqnarray}
Here, $\beta(\textbf{x})$ is a scalar function that depends upon the choice of curvilinear coordinates.

However, since the state vector can be expanded as $\mathbf{x}=p(\mathbf{x})\mathbf{p}+\phi(\mathbf{x})\mathbf{\hat\phi}$ where $\mathbf{p}\equiv\nabla_{\mathbf{x}}
E_{0}/|\nabla_{\mathbf{x}}E_{0}|$ and $\mathbf{\hat\phi}\equiv\mathbf{n}\times\mathbf{p}$, $E_{0}$ must be only a function of
$p(\mathbf{x})$. Moreover, if the energy  is independent of time, one can take $E_{0}(\mathbf{x})$ itself as a canonical momentum, that is,
$p\equiv E_{0}(\mathbf{x})$. Then in terms of $(p,\phi)$, Eq. (\ref{equaforx}) will take a canonical form
\setlength\abovedisplayskip{6pt}
\setlength\belowdisplayskip{6pt}
\begin{eqnarray}
\dot{p}&=&0\equiv-\frac{\partial H_{O}}{\partial \phi},\nonumber\\
\dot{\phi}&=&\frac{2\pi}{T(E_{0})}\equiv\frac{\partial H_{O}}{\partial p},
\label{cycliccoord}
\end{eqnarray}
where $H_{O}(p)=\int dp'\dot{\phi}(p')$ is the Hamiltonian in the old frame. $T(E_{0})$ is the period of the dynamic state trajectories $C(E_{0})$, which can be calculated from Eq. (\ref{equaforx}) as
\setlength\abovedisplayskip{6pt}
\setlength\belowdisplayskip{6pt}
\begin{eqnarray}
T(E_{0})&\equiv&\oint_{C(E_{0})}\frac{dl}{v(\mathbf{x})},\nonumber\\
&=&\oint_{C(E_{0})}\frac{dy}{\dot{y}}=\oint_{C(E_{0})}\frac{dx}{\dot{x}}.
\label{Period}
\end{eqnarray}
Here, $dx$ and $dy$ are the components of the displacement element $dl$ along the trajectories $C(E_{0})$ that are projected on the x and y axes, respectively.  Notably, from the form of Eq. (\ref{cycliccoord}), $\phi$ is a \textit{cyclic} coordinate in terms of the canonical formalism\cite{goldstein2014classical,Taniguchi2014b}.

In terms of $(p,\phi)$, the exact energy balanced equation can be expressed as
\setlength\abovedisplayskip{6pt}
\setlength\belowdisplayskip{6pt}
\begin{eqnarray}
\frac{d E_{0}}{dt} = \dot{p}(\mathbf{x}).
\end{eqnarray}
Here, it should be noted that because the energy dissipation rate generally depends not only on $p$ but also on $\phi$, as can be seen from Eq. (\ref{vectformappro}), the time rate of $p$ will probably be dependent on $\phi$. Thus, we have to take an average of $\dot{p}$ during one period of $T(E_{0})$ to obtain an averaged $\dot{p}$ over $\phi$, e.g. IP-STNO\cite{kiselev2003microwave} and PMA-STNO \cite{Kubota2013}.
In regard to the non-conservative part, we therefore have
\setlength\abovedisplayskip{6pt}
\setlength\belowdisplayskip{6pt}
\begin{eqnarray}
\dot{p}_{\mathrm{av}}(p) &=&\left\langle\frac{d p}{d t}\right\rangle_{T\left(E_{0}\right)},\nonumber \\
&=&\left[\frac{1}{T\left(E_{0}\right)} \int_{0}^{T} d t \frac{dp}{dt}\right],\nonumber \\ &=&\left[\frac{1}{T\left(E_{0}\right)} \oint_{C\left(E_{0}\right)} \left(\frac{d y}{\dot{y}}\right) \dot{p}(\mathbf{x})\right],\nonumber\\
&\equiv&-\alpha S(p)\left(\frac{\partial H_{O}}{\partial p}\right)+a(p,\mu).
\end{eqnarray}
Here, the first term on the right-hand side of the last equality is expressed by the general form of the \textit{Rayleigh dissipation} with a damping constant $\alpha$ and a positive damping function $S(p)$, which measures the damping rate. The second term is the negative damping function. Note that the above approach is valid only when the energy injected by the non-conservative part during a single period of the conserved trajectory is significantly smaller than the energy level of the trajectory, i.e. $|\Delta E|=|\int_{0}^{T(E_{0})} dt\dot{p}|<E_{0}$.

 Eq. (\ref{vectform}) can subsequently be approximately expressed as
\setlength\abovedisplayskip{6pt}
\setlength\belowdisplayskip{6pt}
\begin{eqnarray}
\dot{p}&\approx&-\alpha S(p)\left[\frac{\partial H_{O}}{\partial p}-\frac{a(p,\mu)}{\alpha S(p)}\right],\nonumber\\
\dot{\phi}&=&\frac{\partial H_{O}}{\partial p}.
\label{apppphi}
\end{eqnarray}
Next, by using the local coordinate transformation (see Eq. (\ref{RotTransgeneral})), we have
\setlength\abovedisplayskip{6pt}
\setlength\belowdisplayskip{6pt}
\begin{eqnarray}
(p)_{\mathrm{O}}&=&(p)_{\mathrm{N}}=p,\nonumber\\
\phi&=&\Phi+v_{p}(p)t.\nonumber
\end{eqnarray}
Eq. (\ref{apppphi}) in the new frame has the following form:
\setlength\abovedisplayskip{6pt}
\setlength\belowdisplayskip{6pt}
\begin{eqnarray}
\dot{p}&=&-\alpha S(p)\frac{\partial H_{N}}{\partial p}
,\nonumber\\
\dot{\Phi}&=&\frac{\partial H_{N}}{\partial p}.
\label{appnonaxiaequa}
\end{eqnarray}
Here the Hamiltonian $H_{N}$ in the new frame is given as
\setlength\abovedisplayskip{6pt}
\setlength\belowdisplayskip{6pt}
\begin{eqnarray}
H_{N}(p)=H_{O}(p)-\int^{p} dp'v_{p}(p').\nonumber
\end{eqnarray}
Here, $v_{p}(p)=a(p,\mu)/[\alpha S(p)]$. By requiring $(\partial H_{N}/\partial p)_{p_{0}}=0$ as well as $(\partial^{2} H_{N}/\partial p^{2})_{p_{0}}>0$, i.e.
\setlength\abovedisplayskip{6pt}
\setlength\belowdisplayskip{6pt}
\begin{eqnarray}
H_{O}^{(1)}(p_{0})&=&\frac{a(p_{0},\mu)}{\alpha S(p_{0})},\nonumber\\
H_{O}^{(2)}(p_{0})&>&v^{(1)}_{p}(p_{0}),
\label{solvep0}
\end{eqnarray}
one can easily analyze the stability of equilibrium points $p_{0}$, which indicate the stable oscillations in the old frame.\\

\subsection{\label{app:2}Generalized Pendulum-like Model}

For the case of multiple auto-oscillators with interactions, Eq. (\ref{appnonaxiaequa}) can be extended to the following form:
\setlength\abovedisplayskip{6pt}
\setlength\belowdisplayskip{6pt}
\begin{eqnarray}
\dot{p_{i}}&\approx&-\alpha_{i} S_{i}(p_{i})\frac{\partial H_{N,i0}}{\partial p_{i}}-\frac{\partial H_{N}}{\partial \Phi_{i}},\nonumber\\
\dot{\Phi}_{i}&=&\frac{\partial H_{N}}{\partial p_{i}}.
\label{appnonaxiaequagen}
\end{eqnarray}
Here, the total Hamiltonian $H_{N}$ is
\setlength\abovedisplayskip{6pt}
\setlength\belowdisplayskip{6pt}
\begin{eqnarray}
H_{N}(p,\Phi,t)&=&\sum_{i=1}^{n}H_{Ni,0}(p_{i})\nonumber\\
&&+\frac{1}{2}
\sum_{i,j=1(i\neq j)}^{n}U'_{I}(p_{i},p_{j},\Phi_{i},\Phi_{j},t),\nonumber\\
&=&\sum_{i=1}^{n}\bigg[H_{Oi}(p_{i})+U'_{Ni}(p_{i})
\bigg]\nonumber\\
&&+\frac{1}{2}
\sum_{i,j=1(i\neq j)}^{n}U'_{I}(p_{i},p_{j},\Phi_{i},\Phi_{j},t).\nonumber\\
\end{eqnarray}
Here, $U'_{Ni}(p_{i},t)=-\int^{p_{i}} dp'v_{pi}(p'_{i})$ is the effective potential induced by the local coordinate transformations applied to each individual oscillator. $U'_{I}$ is the anisotropic coupling among the oscillators, which satisfy $|(1/2)\sum_{i,j=1(i\neq j)}^{n}U'_{I}|\ll|H_{Ni,0}|$. This means that the coupling potentials will not significantly distort the dynamic state trajectories of each oscillator. Therefore, the stable oscillatory states still satisfy the requirements that $\left(\partial H_{Ni,0}/\partial p_{i}\right)_{p_{i0}}=0$ and $\left(\partial^{2} H_{Ni,0}/\partial p_{i}^{2}\right)_{p_{i0}}>0$.\\

Around these stable states, the Hamiltonian can be approximated as
\setlength\abovedisplayskip{6pt}
\setlength\belowdisplayskip{6pt}
\begin{eqnarray}
H_{N}(p,\Phi,t)&\approx&
\frac{1}{2}\sum^{n}_{i=1}H^{(2)}_{N,i0}(p_{i0})
\delta p_{i}^{2}\nonumber\\
&&+\frac{1}{2}\sum^{n}_{i,j=1(i\neq j)}U'_{I}(p_{i},p_{j},\Phi_{i},\Phi_{j},t).\nonumber\\
\label{ApproHN}
\end{eqnarray}
Here, $\delta p_{i}\equiv p_{i}-p_{i0}$ is a small deviation away from the equilibrium points $p_{i0}$, and the superscript $(2)$ denotes the second derivative. Since $\mid (1/2)\sum_{i,j=1(i\neq j)}^{n}U'_{I}\mid\ll H^{(2)}_{N,i0}(p_{i0})$, the order of stable state Hamiltonian $\mid H_{N}\mid$ should be around $\mid(1/2)\sum_{i,j=1(i\neq j)}^{n}U'_{I}\mid$ at $p_{i}= p_{i0}+\delta p_{i}$. According to the energy conservation law, we know that the conservative trajectories around $p_{i}=p_{i0}$ given by Eq. (\ref{ApproHN}) satisfy
\setlength\abovedisplayskip{6pt}
\setlength\belowdisplayskip{6pt}
\begin{eqnarray}
H_{N}(p_{0},\Phi_{\mathrm{a}},t_{\mathrm{a}})=
H_{N}(p_{0}+\delta p,\Phi_{\mathrm{b}},t_{\mathrm{b}}),\nonumber
\end{eqnarray}
where the subscripts a and b denote the initial and final states, respectively. The order of
the deviation $\mid\delta p_{i}\mid$ induced by the perturbation of $U'_{I}$, can be easily estimated as
\setlength\abovedisplayskip{6pt}
\setlength\belowdisplayskip{6pt}
\begin{eqnarray}
\mid\delta p_{i}\mid\sim\left[\frac{\mid \Delta U'_{I}\mid}{H^{(2)}_{N,i0}(p_{i0})}\right]^{1/2},
\label{Orderdelp}
\end{eqnarray}
where $\Delta U'_{I}\equiv (1/2) \sum_{i,j,=1(i\neq j)}^{n}U'_{I}(p_{i0},p_{j0},\Phi_{i\mathrm{a}},\Phi_{j\mathrm{a}}
$\\$,\tau_{\mathrm{a}})-U'_{I}(p_{i0}+\delta p_{i},p_{j0}+\delta p_{j},\Phi_{i\mathrm{b}}
,\Phi_{j\mathrm{b}},\tau_{\mathrm{b}})$. For sufficiently small $|\delta p_{i}|$,
 Eq. (\ref{appnonaxiaequagen}) can be reasonably expanded as
\setlength\abovedisplayskip{6pt}
\setlength\belowdisplayskip{6pt}
\begin{eqnarray}
\dot{\delta p_{i}}&\approx&-\alpha_{i} S_{i}(p_{i0})H^{(2)}_{N,i0}(p_{i0})\delta p_{i}-\sum_{j=1(j\neq i)}^{n}\bigg(\frac{\partial U'_{I}}{\partial\Phi_{i}}\bigg)_{p_{i0}},\nonumber\\
\dot{\Phi}_{i}&\approx& H^{(2)}_{N,i0}(p_{i0})\delta p_{i}+\sum_{j=1(j\neq i)}^{n}\bigg(\frac{\partial U'_{I}}{\partial p_{i}}\bigg)_{p_{i0}}.
\label{appnonaxiaequagena}
\end{eqnarray}

Using the transformations,
\setlength\abovedisplayskip{6pt}
\setlength\belowdisplayskip{6pt}
\begin{eqnarray}
(\delta p_{i})_{\mathrm{O}}&=&(\delta p_{i})_{\mathrm{N}}=\delta p_{i},\nonumber\\
(p_{i0})_{\mathrm{O}}&=&(p_{i0})_{\mathrm{N}}=p_{i0},\nonumber\\
\phi_{i}&\approx&\Phi_{i}+[v_{pi}(p_{i0})+v^{(1)}_{pi}(p_{i0})\delta p_{i}]t,\nonumber\\
H^{(2)}_{N,i0}(p_{i0})&=&H^{(2)}_{Oi}(p_{i0})-v^{(1)}_{pi}(p_{i0}),\nonumber
\end{eqnarray}

we easily obtain the old frame version of Eq. (\ref{appnonaxiaequagena}):
\setlength\abovedisplayskip{6pt}
\setlength\belowdisplayskip{6pt}
\begin{eqnarray}
\dot{\delta p_{i}}&\approx&-\alpha_{i} S_{i}(p_{i0})H^{(2)}_{N,i0}(p_{i0})\delta p_{i}-\sum_{j=1(j\neq i)}^{n}\bigg(\frac{\partial U_{I}}{\partial\phi_{i}}\bigg)_{p_{i0}},\nonumber\\
\dot{\phi}_{i}&\approx&v_{pi}(p_{i0})+H^{(2)}_{Oi}(p_{i0})\delta p_{i}+\sum_{j=1(j\neq i)}^{n}\bigg(\frac{\partial U_{I}}{\partial p_{i}}\bigg)_{p_{i0}}.\nonumber\\
\label{appnonaxiaequagenaold}
\end{eqnarray}
Notably, due to the presence of $H^{(2)}_{Oi}(p_{i0})$, the dynamics of phase angle $\phi_{i}$ will be coupled with that of  momentum $\delta p_{i}$. The role of the coefficient $H^{(2)}_{Oi}(p_{i0})$ is similar to the \textit{nonlinear frequency shift coefficient} defined in the universal model\cite{Slavin2009}. If we
 make a transformation $\psi_{i}=\phi_{i}-v_{pi}(p_{i0})t$, then the phase angle equation in Eq. (\ref{appnonaxiaequagenaold}) becomes
\setlength\abovedisplayskip{6pt}
\setlength\belowdisplayskip{6pt}
\begin{eqnarray}
[H^{(2)}_{Oi}(p_{i0})]^{-1}\dot{\psi}_{i}&\approx&\delta p_{i}+[H^{(2)}_{Oi}(p_{i0})]^{-1}\sum_{j=1(j\neq i)}^{n}\bigg(\frac{\partial U_{I}}{\partial p_{i}}\bigg)_{p_{i0}}.\nonumber
\end{eqnarray}
Interestingly, this equation resembles a Newtonian particle with $p=mv$. That is,
the effective mass can be defined here as $m_{\mathrm{eff},i}(p_{i0})\equiv|[H^{(2)}_{Oi}(p_{i0})]^{-1}|$. This implies that the particle
with less mass (stronger $H^{(2)}_{Oi}(p_{i0})$) will be more sensitive to time change of momentum $\dot{\delta p_{i}}$, i.e. a larger angular acceleration $\ddot{\psi}$ or $\ddot{\phi}$. Conversely, for a very small $H^{(2)}_{Oi}(p_{i0})$, $\ddot{\psi}$ or $\ddot{\phi}$ can be reasonably neglected for the same order of $\dot{\delta p_{i}}$ due to the huge mass of the particle, so the dynamics of phase angle $\phi_{i}$ will be decoupled with that of $\delta p_{i}$ and governed by Alder's equation\cite{Adler1973,Slavin2009,Zhou2010}.

Therefore, for the general case with stronger $H^{(2)}_{Oi}(p_{i0})$, one can easily obtain the generalized pendulum-like equation by taking the time derivative of both sides of $\phi_{i}$ equation of Eq. (\ref{appnonaxiaequagenaold}):
\setlength\abovedisplayskip{6pt}
\setlength\belowdisplayskip{6pt}
\begin{eqnarray}
\ddot{\phi}_{i}&=&-\left[\alpha_{i}S_{i}(p_{i0})H^{(2)}_{N,i0}(p_{i0})\right]\dot{\phi}_{i}
+H^{(2)}_{N,i0}(p_{i0})a_{i}(p_{i0},\mu_{i})\nonumber\\
&&+\alpha_{i}S_{i}(p_{i0})H^{(2)}_{N,i0}(p_{i0})\sum_{j=1(j\neq i)}^{n}\left(\frac{\partial U_{I}}{\partial p_{i}}\right)_{p_{i0}}-H^{(2)}_{Oi}(p_{i0})
\nonumber\\
&&\times\sum_{j=1(j\neq i)}^{n}\left(\frac{\partial U_{I}}{\partial\phi_{i}}\right)_{p_{i0}}
+\sum_{l=1}^{n}\frac{\partial}{\partial\phi_{l}}\left(\frac{\partial U_{I}}{\partial p_{i}}\right)_{p_{i0}}\dot{\phi}_{l}.
\label{nonconpenduN}
\end{eqnarray}
Note that the term related to $\dot{\delta p_{i}}$ has been replaced with the $\delta p_{i}$ equation from Eq. (\ref{appnonaxiaequagenaold}).

\nocite{*}

\bibliography{HH_phaselocking}

\end{document}